\documentclass[journal]{IEEEtran}
\usepackage{graphicx,amssymb,lineno}
\usepackage{amsmath,amsfonts,amssymb}

\newtheorem{proposition}{Proposition}
\newtheorem{definition}{Definition}
\usepackage{algorithm}
\usepackage{algorithmic}
\usepackage[usenames]{color}
\usepackage{float}
\usepackage{epstopdf}
\usepackage{longtable,booktabs}
\usepackage[numbers,sort&compress]{natbib}
\usepackage{mathtools}
\usepackage[justification=centering]{caption}

\usepackage{graphicx,graphics,color,epsfig,subfigure,graphpap,rotate}
\usepackage{times, verbatim, subfigure, epsfig, graphicx, latexsym, amsmath}
\usepackage{url}
\usepackage{subfigure}
\begin{document}
%
\title{Sub-Channel Allocation for Device-to-Device Underlaying Full-Duplex mmWave Small Cells using Coalition Formation Games
}

\author{Yibing~Wang,
        Yong~Niu,~\IEEEmembership{Member,~IEEE},
        Hao~Wu,
        Zhu~Han,~\IEEEmembership{Fellow,~IEEE},
        Bo~Ai,~\IEEEmembership{SeniorMember,~IEEE},
        and Qi~Wang%

\thanks{Copyright (c) 2015 IEEE. Personal use of this material is permitted. However, permission to use this material for any other purposes must be obtained from the IEEE by sending a request to pubs-permissions@ieee.org.}

\thanks{Y. Wang, Y. Niu, H. Wu, and B. Ai are with the State Key Laboratory of Rail Traffic
 Control and Safety, Beijing Jiaotong University, Beijing 100044, China (E-mails:
 18111034@bjtu.edu.cn, niuy11@163.com).} %

\thanks{
Z. Han
is with the University of Houston, Houston, TX 77004 USA (E-mail:
zhan2@uh.edu), and also with the Department of Computer Science and Engineering,
Kyung Hee University, Seoul, South Korea, 446-701.  } %

\thanks{
Q. Wang
is with Huawei Device Co. Ltd,
Shenzhen 518129, China (E-mail:
steven\_wq@hotmail.com).}

\thanks{This study was supported by National Key R\&D Program of China under Grant 2016YFE0200900; and by the National Natural Science Foundation of China Grants 61725101, 61801016, 61971030, and U1834210; and by the China Postdoctoral Science Foundation under Grant 2017M610040 and 2018T110041; and by the Beijing Natural Fund under Grant L172020; and by Major projects of Beijing Municipal Science and Technology Commission under Grant No. Z181100003218010; in part by the State Key Lab of Rail Traffic Control and Safety, Beijing Jiaotong University, under Grants RCS2019ZZ005, 2017RC031, and RCS2019ZT006; in part by US MURI AFOSR MURI 18RT0073, NSF EARS-1839818, CNS1717454, CNS-1731424, CNS-1702850, and CNS-1646607.}

}

\maketitle

\begin{abstract}
The small cells in millimeter wave (mmWave) have been utilized extensively for the fifth generation (5G) mobile networks. And full-duplex (FD) communications become possible with the development of self-interference (SI) cancellation technology. In this paper, we focus on the optimal allocation of the sub-channels in the small cells deployed densely underlying the mmWave networks. In order to improve the resource utilization and network capacity, device-to-device (D2D) communications are adopted in networks. For maximizing the system throughput, we propose a cooperative sub-channel allocation approach based on coalition formation game, and theoretically prove the convergence of the proposed allocation algorithm. The utility of the coalition formation game is closely related to the sum system throughput in this paper, and each link makes an individual decision to leave or join any coalition based on the principle of maximizing the utility. Besides, a threshold of the minimum transmission rate is given to improve the system fairness. Through extensive simulations, we demonstrate that our proposed scheme has a near-optimal performance on sum throughput. In addition, we verify the low complexity of the proposed scheme by simulating the number of switch operations.

\end{abstract}

\section{Introduction}\label{S1}
In the fifth generation (5G) mobile cellular network, millimeter wave (mmWave) bands have been proposed to support various multi-gigabit wireless services. For the challenge of the explosive mobile traffic growth, the mmWave is a promising candidate in all respects. In the mmWave networks, directional antenna and beamforming technique are common means of compensating for the non-negligible path loss \cite{MIMO1, MIMO2, Beamforming, Y1}. Besides, concurrent transmission can significantly increase the system throughput \cite{mao3}. Due to the small wavelength of the mmWave, the antenna arrays composed of directional antennas are able to be synthesized in some small platforms \cite{CMOS}, and then the beams of the transmitter and receiver point to each other by beam training \cite{beam_training}. However, the effective range of mmWave communication is limited by the directional antenna. Thus, small cells under the mmWave bands have been proposed to solve the problem of limited transmission range, as well as obtaining the gains from high bandwidth and spatial reuse.

In addition to the mmWave directional communication and small cell, device-to-device (D2D) communications are always used to share the spectrum resource with access users \cite{D2D2}. D2D communications can reduce the large path loss caused by long distance \cite{D2D1}, and support massive content-based services, some related specific applications are like advertising message distribution, broadcasting services and so on \cite{Coalitional4, Leveraging5, D2D3, Y2}. In addition, D2D communications can also cover the shaded area of mmWave communications and solve the blockage problem. To further improve the transmission efficiency, full-duplex (FD) communication can be considered, since FD communication allows the equipments to transmit and receive simultaneously in the same channel. Compared with half-duplex (HD), the spectral efficiency can be theoretically doubled by FD communication \cite{15,16,17}. Due to the restriction of self-interference (SI), FD communication was not used widely before. However, with the continuous development of the SI cancellation technology, FD communication has become a research focus recently \cite{10}. Although SI cancellation technology is evolving constantly, there is still some residual self interference (RSI) in the system, and SI can not be completely eliminated in practice. Therefore, the interferences should be considered are not only multiuser interference (MUI), but also the RSI from FD communications.

In this paper, we consider a scenario of D2D communications in the small cells underlying the mmWave networks. Base stations (BSs) and mobile user devices are all equipped with directional antennas and work in FD mode. Compared to HD communications in traditional cellular networks, directional communications and FD communications in mmWave networks have significantly improved the performance on throughput. However, the impact of channel allocation on network performances also cannot be ignored. In this case, we research on how to allocate sub-channels to transmission links for the maximum system throughput. In the channel allocation problem, the coordinations among transmission links seriously affect the transmission performance, and coalition formation game is exactly good at making the game participants (which are the links in the paper) cooperate as fully as possible to obtain the maximum utility \cite{gameh, ex10}. Therefore, we propose a novel cooperation approach based the coalition formation game which is different from other existing channel allocation approaches. We model the sub-channel allocation problem as a coalitional game to maximize the revenues obtained from the transmission links. The links in the game can select to occupy different sub-channels with cooperation, and the sub-channel allocation concretely depends on the resource-sharing possibilities of the different links in the network \cite{Coalitional4, Games, Games2, Games3, Games4}. For the coalition formation algorithm of the proposed approach, the links can eventually self-organize into a Nash-stable partition to maximize the throughput of all the links. Compared with other non-cooperative approaches, the proposed approach based the coalition formation game can achieve the near-optimal performance with the relatively low complexity, and coordinate the entire network from a macro perspective. For the sub-channel allocation algorithm  based coalition formation game, we fully combine the advantages of FD communications and D2D communications to maximize the transmission efficiency in the entire system. Thus, the contributions of this paper can be summarized as follows
\begin{itemize}
\item We make use of the advantages of FD communications and D2D communications for channel allocation in the mmWave small cells. Compared with the conventional communication schemes in traditional cellular networks, D2D communications in mmWave bands have less long distance path loss and can cover the shaded area. In addition, FD communications are adopted to make more links concurrent transmit so that the system throughput and transmission efficiency can increase obviously.
\item We model the channel allocation problem as a coalition formation game and propose a corresponding coalition formation algorithm for forming the coalitions. The links of the coalition formation game can self-organize a Nash-stable final partition in terms of the system throughput maximization. Besides, we specify the minimum transmission rate to prevent links from having unfairness with extremely low transmission rates in proposed coalition cooperation approach.
\item The proposed approach is compared with other sub-channel allocation strategies by extensive simulations, and the evaluation results show that the proposed algorithm for sub-channel allocation can significantly increase the system throughput and accomplish all the link transmissions with better fairness. Moreover, we also demonstrate the low complexity of our algorithm via the average number of switch operations.

\end{itemize}

The rest of the paper is organized as follows. Section \ref{S2} introduces the related work. Section \ref{S3} introduces the system overview and illustrates the system model. Section~\ref{S4} formulates the optimal sub-channel allocation problem for links in mmWave small cells. In Section~\ref{S5}, we model the sub-channel allocation problem as a coalitional game, and propose a relevant coalition formation algorithm. Section \ref{S6} shows the performance evaluations of the scheme we proposed. In Section \ref{S7}, we conclude this paper.

\section{Related Work}\label{S2}
Recently, there have been many works on the mmWave research, such as \cite{ex12, ex13, ex14, ex16, ex17}. In \cite{ex12}, a joint scheduling scheme is proposed for improving the performances of mmWave networks. In \cite{ex13}, the conditions of concurrent transmission exclusive region are deduced, and the REX scheduling scheme is proposed to acquire significant spatial reuse gain. In \cite{ex14}, a concurrent transmission scheduling scheme is proposed to make more links transmit concurrently with quality of service (QoS) requirements satisfied. In \cite{ex16}, the authors focused on the problem of blockage traffic, and exploited multi-hop paths to relay the blocked flows. In \cite{ex17}, the frame based directional MAC protocol is proposed, and a greedy coloring (GC) algorithm is presented for link scheduling of the concurrent transmissions. However, these works are all based on HD communications, and have no relationship with D2D communications and channel allocation.

There also have been some works about channel allocation. In \cite{chan1}, a distributed protocol for channel allocation in wireless sensor networks is proposed with theoretical bounds, but the objective of this protocol is to minimize the maximum interference between links. In \cite{D2D4}, a framework based on stochastic geometry is proposed to obtain the optimal channel allocation and achieve the superior network performance efficiently, yet the channel allocation is only for the uplink. In \cite{chan2}, the authors proposed a joint power and channel allocation algorithm based the Hungarian algorithm, and the power allocation is derived independently on each sub-channel. The joint power and channel allocation is independent without user cooperation and the research is in the traditional cellular networks. In \cite{chan3}, the authors presented a natural non-cooperative game and the utility is the sum rate of all the players. This approach is based on game, but is non-cooperative.

In addition, the research on cooperation schemes based on coalition formation game has yielded certain results, for instance, \cite{Coalitional4, ex11, ex4, ex5, ex7, ex8, ex9, ex10}. In \cite{Coalitional4}, the authors focused on the problem of uplink resource allocation for both D2D and cellular users, and a coalition formation game based approach is proposed to obtain the maximum system sum rate. In \cite{ex11}, a two-step coalition game is proposed to model the uplink D2D resource allocation problem, and found out the efficient allocation scheme to share the cellular uplink channels eventually. In \cite{ex5}, the authors addressed the downlink resource allocation problem for the users in D2D and cellular communications, by proposing a reverse iterative based combinatorial auction approach. However, these three approaches only consider the resource allocation of the uplink resource or downlink resource. In \cite{ex9}, the social relationships of the resource allocation of D2D communications are introduced, then a social group utility maximization game is formulated for the maximum social group utility of each user in D2D communications. In \cite{ex10}, the coalition game is used to model the distribution problem of the dynamic popular content in vehicular networks, and the approach is proposed, which on-board units exchange their possessed partial content with their neighbors by broadcasting and receiving to obtain the complete content. Ref. \cite{ex10} and \cite{ex9} only propose the schemes for the resource allocation of links between devices or vehicles, yet they do not consider resource allocation of access links. Besides, all these works are based on HD communications in the cellular networks, so the performance in terms of the system throughput is subject to further improvement.

Thus, for the channel allocation of both access links and D2D links in mmWave networks, the FD cooperation approach based on coalition formation game is need to be proposed to achieve the superior performance on throughput.

\begin{figure*}[t!]
  \begin{center}
  \includegraphics[width=10cm]{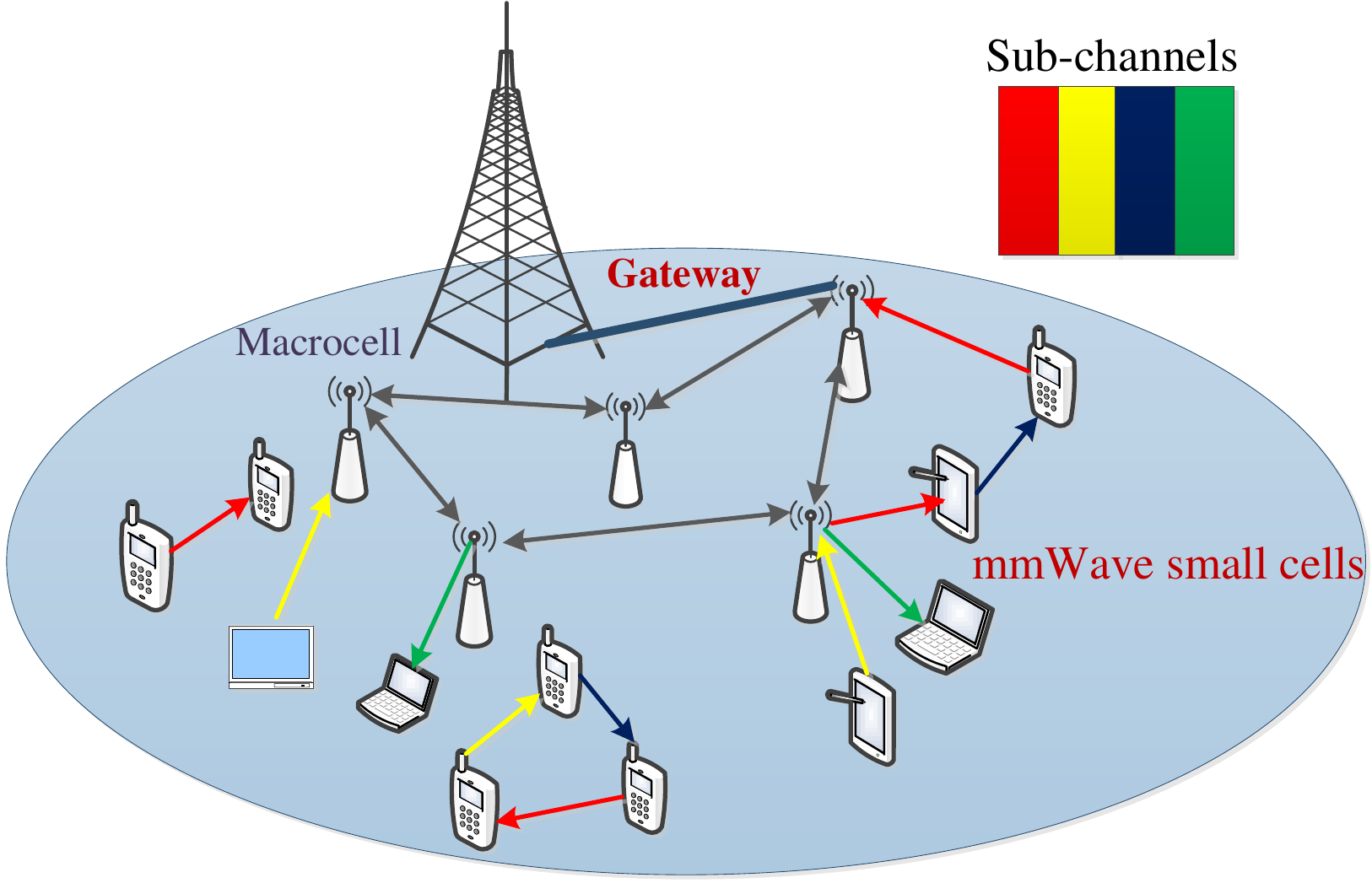}
  \caption{An example of sub-channel allocation for mmWave small cells densely deployed underlying the macrocell network.} \label{fig:example}
  \end{center}
\end{figure*}

\section{System Overview}\label{S3}


\subsection{System Model}\label{S3-1}

\begin{figure}[t!]
  \includegraphics[width=8cm]{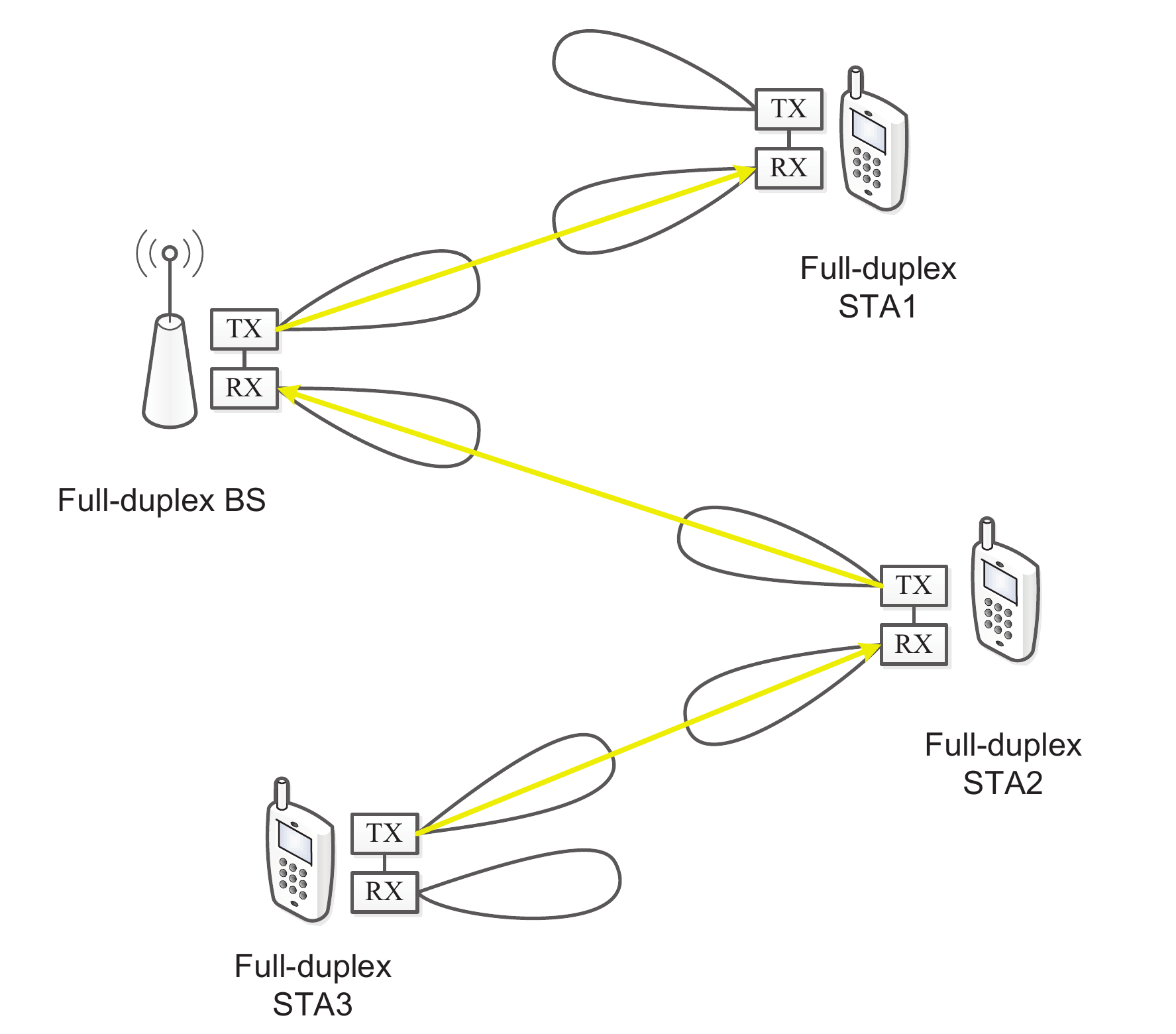}
  \caption{Full-duplex communication in a mmWave small cell.} \label{fig:FD_node}
\end{figure}

In this paper, we consider a mmWave network under the scenario of small cells densely deployed, as shown in Fig. \ref{fig:example}. We consider there are $N$ base stations (BSs) in network, and each mobile user equipment may transmit in D2D mode or connect to the nearest BS in cellular mode \cite{D2D1}. The BSs connected to the backbone network via the macrocell are called gateways. There are amount of traffic flows, which are from BS to mobile device or from one device to another device, so we consider both access links and D2D links. The mmWave bands can be divided into multiple available sub-channels, as well as the access links in different sub-channels are able to be scheduled simultaneously. The low-cost and compact directional antennas are now crucial and feasible in mmWave bands \cite{beam_training, ant4}. To simplify the channel allocation problem, we assume each user device is equipped with two directional antennas for transmitting and receiving data at the same time, as shown in Fig. \ref{fig:FD_node}. Multiple sub-channels in the mmWave bands are shared by both access links and D2D links equally. In the above scenario, we focus on the sub-channel allocation problem to gain the maximum network capacity. In order to facilitate the description, both BSs and mobile users in the network are treated as nodes in the following studies.



For link $i$, we denote its transmitter by $s_i$, and denote its receiver by $r_i$. Node $s_i$ and node $r_i$ point towards each other by their directional beams for directional transmissions. The received signal power at $r_i$ from node $s_i$ can be calculated as \cite{ex13}
\begin{equation}\label{eqp}
{P_r}(i,i) = {k_0}G^t_{s_i}G^r_{r_i}l_{s_ir_i}^{ - n}{P_t},
\end{equation}
where ${k_0}$ is a constant coefficient and proportional to ${(\frac{\lambda }{{4\pi }})^2}$ ($\lambda $ denotes the wavelength), $n$ is the path loss exponent, and ${P_t}$ is the transmission power \cite{ex13}. $G^t_{s_i}$ denotes the gain of transmitter antenna at node $s_i$, and $G^r_{r_i}$ denotes the gain of receiver antenna at node $r_i$. $l_{s_ir_i}$ is the length of the link from $s_i$ to $r_i$.

Under FD transmissions, the interferences between links can be classified into two cases, RSI and MUI. For any two concurrent links in the same sub-channel, when the receiver of one link acts as the transmitter of the other one, the first case RSI is the residual SI after performing SI cancellation technology.
The RSI of the link $i$
can be modeled in terms of the transmission power of another co-channel link, which transmits from the receiver of link $i$.
Therefore, we can use $\beta_{r_i} P_{t^\prime}$ to denote the RSI of link $i$, where the non-negative parameter $\beta_{r_i}$ represents the SI cancellation level at node $r_i$. $P_{t^\prime}$ is the transmission power of the link which transmits from node $r_i$. For link $i$, the co-channel concurrent link from node $r_i$ can only have one at most. Thus, the signal to noise ratio (SNR) of link $i$ at node $r_i$ can be calculated as
\begin{equation}
SNR = \frac{{P_r}(i,i)}{N_0W+\beta_{r_i} P_{t^\prime}},
\end{equation}
where ${P_r}(i,i)$ denotes the received signal power of link $i$, and the specific calculation is defined in (\ref{eqp}). $N_0$ denotes the onesided power spectral density of white Gaussian noise. $W$ is the channel bandwidth.

The other kind of interference MUI is the interference between any two co-channel links without sharing the same node.
For another link $j$, if it is a concurrent link of link $i$ in the same sub-channel, the interference at $r_i$ from node $s_j$ can be calculated as
\begin{equation}
{I_{ji}} = \rho{k_0}G^t_{s_j}G^r_{r_i}l_{s_jr_i}^{ - n}{P_t},
\end{equation}
where $\rho$ is the factor of MUI, and the specific value is related to the signals cross correlation from different co-channel links \cite{ex14}.

There is a lack of the multipath effect, so Gaussian channel can be adopted to model the channel of the directional mmWave links  \cite{Y2}.
Considered MUI and RSI of the simultaneous transmission links, we can obtain the transmission rate of link $i$ according to Shannon's channel capacity
as
\begin{equation}
R_i=\eta W{\rm{log}}_2\left(1+\frac{{P_r}(i,i)}{N_0W+\beta_{r_i} P_{t^\prime}+\sum\limits_{j}I_{ji}}\right),\label{equation:rate}
\end{equation}
where $\eta$ describes the efficiency of the transceiver design in the range of $(0,1)$. Due to the certain power loss at the receiver and transmitter, setting the parameter $\eta$ represents the loss of this part in the calculation.

\subsection{Antenna Model}\label{S3-2}
In this paper, we assume BSs and devices are equipped with directional antennas. Thus, we adopt the realistic directional antenna model mentioned in \cite{23}, which is the reference antenna model with sidelobe for IEEE 802.15.3c. The gain of a directional antenna in units of dB, which is denoted by $G(\theta)$, can be expressed as
\begin{equation}
G(\theta)=\left\{
\begin{array}{rcl}
&G_0-3.01\cdot (\frac{2\theta}{\theta_{-3dB}})^2, &{0^{\circ}\leq \theta \leq \theta_{ml}/2}, \\
&G_{sl}, &{ \theta_{ml}/2\leq \theta \leq 180^{\circ}},
\end{array}
\right.\label{equation:Antenna Model}
\end{equation}
where $\theta$ denotes an arbitrary angle within the range $[0^{\circ}, 180^{\circ}]$, $\theta_{-3dB}$ denotes the angle of the half-power beamwidth, and $\theta_{ml}$ denotes the main lobe width in units of degrees, and $\theta_{ml}=2.6\cdot\theta_{-3dB}$. $G_0$ is the maximum antenna gain, which is obtained by $G_0 = 10\log(1.6162/\sin(\theta_{-3dB}/2))^2$, $G_{sl}$ is the sidelobe gain, which is expressed as $G_{sl} = -0.4111\log(\theta_{-3dB})-10.579$.


\section{Problem Formulation}\label{S4}

Considering the typical mmWave network, when the distance between any two mobile users is within a certain range, D2D users can communicate directly and establish a D2D link. If the distance is large, regular users can communicate through BSs and establish access links. There are several sub-channels in the system, and each sub-channel can be shared by access links and D2D links equally at the same time.

There is a certain group of links being scheduled concurrently in the same sub-channel. For any two co-channel links, if the transmitter of one link is the receiver of the other one, these two links can transfer simultaneously. Moreover, each BS can be the transmitters or receivers of multiple access links. FD communications can make more links transmit simultaneously, but more co-channel concurrent links may be affected by severe interference and cause the transmission rates of some links to be too low. So a threshold of the minimum transmission rate is denoted by $R_{min}$, and a judgment need to be made by $R_{min}$. Coordinating the links to occupy different sub-channels, $R_{min}$ is able to prevent the transmission rate of any link from being lower than $R_{min}$. In a word, $R_{min}$ can improve the fairness of the links, and avoid unfairness which is due to low transmission rates of some links.

We denote the set of access links by $A$ and the set of D2D links by $D$. Any sub-channel is denoted by $c$ and the set of sub-channels is denoted by $C$. To simplify the representation, we define a binary variable ${a_i^c}$ to indicate whether access link $i$ is scheduled in the sub-channel $c\in C$. If so, ${a_i^c=1}$; otherwise, ${a_i^c=0}$. We further assume that the transmission power of each link is fixed. Let $P_t^i$ indicate the transmission power of access link $i$. Then we can obtain the actual transmission rate of access link $i$ in the sub-channel $c$ as
\begin{equation}
R_i^c=\eta W{\rm{log}}_2\left(1+\frac{a_i^c{P_r}(i,i)}
{\splitfrac
               {N_0W+\sum\limits_{l}a_l^c\beta_{r_i}P_t^l}{+\sum\limits_{u}a_u^cI_{ui}
               +\sum\limits_{w}d_w^cI_{wi}}}\right),\label{equation:actualrate}
\end{equation}
where $l$ denotes the co-channel concurrent link of access link $i$, as well as its transmitter is the receiver of access link $i$, $l\in A\cup D$. Apparently, under our assumption, $l$ cannot exceed 1. $u\in A$ and $w\in D$ denote the access links and D2D links, respectively, which simultaneously occupy the same sub-channel with access link $i$ and have MUI on link $i$. $G_i$ denotes the total gain of link $i$ at node $r_i$.

Similarly, we define a binary variable ${d_j^c}$ to indicate whether D2D link $j$ is scheduled in sub-channel $c$. The actual transmission rate of D2D link $j$ in the sub-channel $c$ is calculated as
\begin{equation}
R_j^c=\eta W{\rm{log}}_2\left(1+\frac
{d_j^c{P_r}(j,j)}
{\splitfrac
          {N_0W+\sum\limits_{l^\prime}d_{l^\prime}^c\beta_{r_j}P_t^{l^\prime}}{+\sum\limits_{u^\prime}a_{u^\prime}^cI_{{u^\prime}j}
          +\sum\limits_{w^\prime}d_{w^\prime}^cI_{{w^\prime}j}}}\right),\label{equation:actualrate}
\end{equation}
where $P_t^j$ indicates the transmission power of D2D link $j$. $G_j$ denotes the total gain of link $j$ at node $r_j$. $l^\prime$ denotes the co-channel concurrent link of D2D link $j$, and its transmitter is the receiver of D2D link $j$. $l^\prime$ also cannot exceed 1. $u^\prime\in A$ and $w^\prime\in D$ denote the access links and D2D links, respectively, which simultaneously occupy the same sub-channel with D2D link $j$ and have MUI on link $j$.

Since each access link or D2D link occupies at most one sub-channel, we have
\begin{align}
\sum\limits_{c\in C}a_i^c\leq 1, \forall i\in A, \label{equation:constraint_1}\\
\sum\limits_{c\in C}d_j^c\leq 1, \forall j\in D. \label{equation:constraint_1-1}
\end{align}

At the same time, the access links with the same transmitters or receivers cannot occupy the same sub-channel, so we have
\begin{equation}
a_i^c+a_{i^\prime}^c\leq 1, \text{ if } s_i=s_{i^\prime} \text{ or } r_i=r_{i^\prime}, \forall i, {i^\prime}\in A, \forall c\in C. \label{equation:constraint_1-2}
\end{equation}

For all the access links which transmit simultaneously, all the uplinks of them have different transmitters and all the downlinks have different receivers. For uplink $i_{up}, i_{up}^\prime\in A$ and downlink $i_{down}, i_{down}^\prime\in A$, the constraint is as follows
\begin{equation}
\begin{split}
s_{i_{up}}\neq s_{i_{up}^\prime}\text{and }r_{i_{down}}\neq r_{i_{down}^\prime}.
\end{split}
\end{equation}

According to the content aforementioned, we can know that any two co-channel D2D links cannot be scheduled simultaneously, unless the transmitter of one link is the receiver of the other link. In other words, the D2D links occupy the same channel can be scheduled simultaneously do not include these links, which have the same transmitters or receivers. Thus, we can obtain the constraint as follows
\begin{equation}
s_{l_d}\neq s_{l_d^\prime}\text{ and } r_{l_d}\neq r_{l_d^\prime}, \forall l_d, l_d^\prime\in D .  \label{equation:constraint_1-3}
\end{equation}

In the system, since the number of sub-channels is limited, the co-channel access links which share the same BS as their transmitters or receivers cannot be too much. Thus, we can obtain the constraint as follows
\begin{equation}
\begin{split}
&\sum\limits_{k=1}^{n}a_{l_k}^c\leq |C|,  l_1, l_2, \cdot\cdot\cdot, l_k\in A, \\ &s_{l_1}=s_{l_2}=\cdot\cdot\cdot=s_{l_k} \text{ or } r_{l_1}=r_{l_2}=\cdot\cdot\cdot=r_{l_k} . \label{equation:constraint_2-2}
\end{split}
\end{equation}


From these formulas and constraints above, we can come to a conclusion, that is, the performances on throughput and fairness are undetermined with uncertain channel allocation scheme. The specific results are closely related to $a_i^c$ and $d_j^c$ although many constraints are given. Thus, the optimal solution of channel allocation cannot be obtain by the gradient descent algorithm, even if $a_i^c$ and $d_j^c$ are relaxed to take the real values. Therefore, a coalitional game approach is proposed in the next section to do the channel allocation efficiently and practically.

\section{Coalitional Game}\label{S5}
In this section, the sub-channel allocation problem is modeled as a coalitional game to acquire the system utility in terms of the system sum rate. The links are regarded as the players of the coalition game. Then we propose a coalition formation algorithm for sub-channel allocation.

\subsection{Coalitional Game }\label{S5-1}
The access links which participate in sub-channel allocation include uplink and downlink. There are $|A|$ access links, $|D|$ D2D links, and $|C|$ sub-channels in the system. The purpose of channel allocation is to achieve a good performance on the system throughput. To devise a coalition formation algorithm for the proposed coalition formation game, we introduce some important definitions of the coalitional game.

\begin{definition}\label{D1}
(A coalitional game) A coalitional game with transferable utility is defined by a pair $(\mathcal{L},U)$, where $\mathcal{L}$ is the set of game players and $U$ is a utility function over the real line for every coalition $S_c\subseteq\mathcal{L}$, and $U(S_c)$ is a real number describing the amount of value that coalition $S_c$ can receive, which can be distributed in any arbitrary manner among the members of coalition $S_c$.
\end{definition}

For the sub-channel allocation, the game players are the links, including D2D links and access links, so we have $\mathcal{L}=A\cup D$. Links occupying the same sub-channel form a coalition, and $S_c$ is the coalition of the links occupying the same sub-channel $c\in C$. There will be more interference between links if there are more links share the sub-channel simultaneously. In this system, access links with the same transmitter or receiver cannot occupy the same sub-channel. Consequently, the links has no motivation to form a grand coalition with occupying only one sub-channel. On the contrary, links prefer to form as many disjoint coalitions of different sub-channels as possible to obtain maximum system throughput. Considering there are $|C|$ sub-channels in the system, the links can be divided into $|C|$ coalitions at most. Therefore, the restrictions of the coalitions are as follows
\begin{equation}
\begin{split}
&\mathcal{L}=S_1\cup S_2\cup\ldots\cup S_{|C|},\\
&S_c\cap S_{c^\prime}=\emptyset, \forall c, c^\prime \in C\;{\rm{and}}\;c\neq {c^\prime}. \label{equation:coalition}
\end{split}
\end{equation}

Considering the RSI and MUI between transmission links, $u$ denotes the access link and $w$ denotes the D2D link in the coalition $S_c$, the set of access links in $S_c$ is $U_c\subseteq A$, and the set of D2D links in $S_c$ is $W_c\subseteq D$, we can obtain the transmission rate of access links $i\in S_c$ as
\begin{equation}
{R_i^c}^\prime=\eta W{\rm{log}}_2\left(1+\frac
{a_i^c{P_r}(i,i)}
{\splitfrac
          {N_0W+\sum\limits_{l}a_l^c\beta_{s_l}P_t^l}{+\sum\limits_{u\in U_c\setminus (i, l)}a_u^cI_{ui}
          +\sum\limits_{w\in W_c}d_w^cI_{wi}}}\right).\label{equation:actualrate1}
\end{equation}

The transmission rate of D2D link $j\in S_c$ can be obtained as
\begin{equation}
{R_j^c}^\prime=\eta W{\rm{log}}_2\left(1+\frac
{d_j^c{P_r}(j,j)}
{\splitfrac
           {N_0W+\sum\limits_{l^\prime}d_{l^\prime}^c\beta_{r_j}P_t^{l^\prime}}{+\sum\limits_{u^\prime\in U_c}a_{u^\prime}^cI_{{u^\prime}j}
           +\sum\limits_{w^\prime\in W_c\setminus (j, l^\prime)}d_{w^\prime}^cI_{{w^\prime}j}}}\right
           ).\label{equation:actualrate1}
\end{equation}

The system throughput at any time is equal to the sum rate of links, so we calculate the sum rate of links in $S_c$ as
\begin{equation}
R(S_c)=\sum\limits_{i\in U_c}{R_i^c}^\prime+\sum\limits_{j\in W_c}{R_j^c}^\prime, \forall c\in C.\label{equation:object1}
\end{equation}

The total revenue of $S_c$ is calculated by the utility function $U(S_c)$. According to the contribution of the transmission links in $S_c$, we define the utility function $U(S_c)$ is proportional to $R(S_c)$, so the utility function
$U(S_c)$ is given by
\begin{equation}\label{eq12}
U(S_c)=\alpha R(S_c)=\alpha\left(\sum\limits_{i\in U_c}{R_i^c}^\prime+\sum\limits_{j\in W_c}{R_j^c}^\prime\right), \forall c\in C,
\end{equation}
where $\alpha>0$ is a utility calculation factor.

According to above content, the definitions of the game formation are included as follows
\begin{itemize}
 \item \emph{Players}: The set of links is denoted as $\mathcal{L}=A\cup D$.
 \item \emph{Coalition}: The players set $\mathcal{L}$ is partitioned into $|C|$ coalitions, $\mathcal{L}=S_1\cup S_2\cup\ldots\cup S_{|C|}$, $S_c\cap S_{c^\prime}=\emptyset$, $\forall c$, $c^\prime \in C\;{\rm{and}}\;c\neq {c^\prime}$.
 \item \emph{Utility}: $U(S_c)$ is the value for each coalition $S_c \subseteq \mathcal{L}$, which is a transferable utility for members in $S_c$, the utility is proportional to the sum rate in this game.
 \item \emph{Strategy}: Players decide to join or leave a coalition based on the utility comparison results of the original coalition and the new coalition.
\end{itemize}

The definitions in this section model the sub-channel allocation problem as a coalitional game with the transferable utility. The utility is proportional to the system throughput, and links tend to form coalitions of different sub-channels to maximize the utility of the coalitional game. Then the algorithm to allocate the sub-channels for all the links is introduced in next section.

\subsection{Coalition Formation Algorithm }\label{S5-2}
First, we give the concept of the coalition partition.

\begin{definition}\label{D0}
(The coalition partition) A coalitional partition is defined as the set $\Pi=\{S_1,\cdot\cdot\cdot,S_k\}$ $(1\leq k\leq|C|)$, which partitions the players set $\mathcal{L}$, i.e., $\forall c, S_c\subseteq\mathcal{L}$ are disjoint coalitions such that $\bigcup_{c=1}^kS_c=\mathcal{L}$.
\end{definition}

In order to maximize the system throughput, preference relation for players to decide whether to join or leave a coalition should be well defined. Instead of initial partition $\Pi=\{S_1,\cdot\cdot\cdot,S_t\}$, a group of players prefers to adopt the utilitarian order to organize themselves into a collection of coalitions $\Pi^\prime=\{S^\prime_1,\cdot\cdot\cdot,S^\prime_{t^\prime}\}$, which is proposed in \cite{Leveraging5, Games2}. Then the utility relationship between two different partitions can be expressed as
\begin{equation}
\sum\limits_{i=1}^{t^\prime}{U(S^\prime_i)}>\sum\limits_{i=1}^{t}{U(S_i)}.\label{equation:compare}
\end{equation}

Therefore, we can get a conclusion that a collection of coalitions $\Pi^\prime$ is very suitable for coalitional games with transferable utility. The definition of the total utility of a collection of coalitions is given as follows
\begin{definition}\label{D3}
(Total utility of coalitions) For a partition $\Pi=\{S_1,\cdot\cdot\cdot,S_k\}$ $(1\leq k\leq|C|)$ of the set $\mathcal{L}$, the total utility can be calculated as
\begin{equation}
U(\Pi)=\sum\limits_{i=1}^k{U(S_i)}.
\end{equation}
\end{definition}

If $U(\Pi^\prime)>U(\Pi)$, the partition $\Pi^\prime$ has a better performance on total utility. Every coalition in $\Pi^\prime$ is the coalition of links which share the same sub-channel. The total utility here is proportional to the sum throughput of the system. Then we give a definition of the preference relation of players in $\mathcal{L}$ as follows
\begin{definition}\label{D4}
(Preference relation $\succ_l$) For any player $l$, a preference relation $\succ_l$ is defined as a complete, reflexive, and transitive binary relation over the set of all coalitions that player $l$ may form.
\end{definition}

For any player $l\in \mathcal{L}$, $S_p\succ_l S_q$ means player $l$ strictly prefers being a member of coalition $S_p$ over being a member of coalition $S_{q}$. So the preference relation of player $l$ between $S_p$ and $S_{q}$ ($S_p\succ_l S_q$) can be quantified as follows
\begin{equation}
\begin{split}
U(S_p\cup l)+U(S_{q}\backslash l)>U(S_p)+U(S_{q}),\\ S_p, S_q\subseteq\mathcal{L}, S_p\neq S_q.\label{equation:quantify}
\end{split}
\end{equation}


Next, we define the coalition switch operation of our coalition game.

\begin{definition}\label{D5}
(Switch Operation) Given a partition $\Pi=\{S_1,\cdot\cdot\cdot,S_k\}$ $(1\leq k\leq|C|)$ of the player set $\mathcal{L}$, if link $l\in\mathcal{L}$ performs a switch operation from $S_p$ to $S_q$ ($S_p, S_q\in\Pi\cup\{\emptyset\}, S_p\neq S_q$), then the  partition $\Pi$ is modified into a new partition as follows
\begin{equation}
\Pi_{new}=(\Pi\backslash\{S_p, S_q\})\cup\{S_p\backslash l\}\cup\{S_q\cup l\}.
\end{equation}
\end{definition}

Then the basic rule for performing switch operations is given as follow

\noindent\emph{\textbf{Switch Rule 1}}: Given a partition $\Pi=\{S_1,\cdot\cdot\cdot,S_k\}$ $(1\leq k\leq|C|)$ of the player set $\mathcal{L}$, for player $\forall l\in\mathcal{L}$, if and only if $S_q\succ_l S_p$ ($S_p, S_q\in\Pi\cup\{\emptyset\}, S_p\neq S_q$), a switch operation from $S_p$ to $S_q$ is allowed.

In other words, for each link $l\in\mathcal{L}$, if the new coalition $\{S_q\cup l\}$ is strictly preferred over its current coalition $S_p$ according to the preference relation defined in (\ref{equation:quantify}), $l$ can leave its current coalition $S_p$ to join another coalition $S_q$ ($S_p, S_q\in\Pi\cup\{\emptyset\}, S_p\neq S_q$).

From Section \ref{S4}, we know that the threshold of the minimum transmission rate $R_{min}$ is needed to make a judgment. It is used to improve the fairness of the proposed cooperation approach, that is, preventing any links from transferring with very low rates. For any link $l$ with the willingness to join the coalition $S_c$, making a judgment on whether the transmission rates of link $l$ and other links in the coalition $S_c$ are bigger than $R_{min}$.

In order to give the rule of the verification switch operation, we assume the current coalition of link $l$ is $S_p$, and $c\in C$ is the sub-channel which is shared by the links in coalition $S_c$. We give a minimum transmission rate $R_{min}$. Then the basic rule for mode switch operations is given as follows

\noindent\emph{\textbf{Switch Rule 2}}: For any link $l$ ($l\in\mathcal{L}$, $\mathcal{L}=A\cup D$), the coalition $S_c$ is selected by $l$ to perform the switch operation from its current coalition $S_p$ with satisfying the preference relation in (\ref{equation:quantify}). At this moment, if link $l$'s participation in coalition $S_c$ makes the rate of link $l$ or any link $l^\prime$ in $S_c$ becomes smaller than $R_{min}$ ($R_l^c<R_{min}$ or $R_{l^\prime}^c<R_{min}$, $\exists l^\prime\in S_c$), the new coalition selected by a switch operation changes from $S_c$ to another coalition (except $S_p$ and $S_c$ ).

In other words, if link $l\in\mathcal{L}$'s participation in a new coalition leads to very small rate of link $l$ or other link in the new coalition, link $l$ should select another sub-channel to obtain a good performance on fairness. The judgment of the rate is made by $R_{min}$.

From these related definitions and switching rules, the pseudo code of coalition formation game for sub-channel allocation is shown in Algorithm \ref{alg: Coalition Formation Algorithm}.

To begin with, there are some preparation work and some parameter initializations. As show in line 1-12, the coalition formation algorithm performs the judgment for the first time to determine whether to perform a switch operation in definition (\ref{D5}). In line 4-10, the first coalition switch operation judgment includes one $R_{min}$ judgement shown in switch rule 2. If the first switch operation is performed with meeting the switch rule 1, the algorithm ends this round of loops and repeats the above operations. If the first switch operation judgment does not meet the switch rule 1, we will further examine the second switch operation in definition (\ref{D5}) under the circumstances that the first switch operation has been performed, which is shown in line 13-33. In line 17-31, there is also one $R_{min}$ judgement in the second coalition switch operation judgment. If the partition after the second switch operation has a higher total utility than the initial partition, these two switch operations will be performed and the current partition will be updated as the new partition after the second switch operation.

\begin{algorithm}[tp!]
\caption {Coalition Formation Algorithm for Sub-channel Allocation} \label{alg: Coalition Formation Algorithm}
\noindent\textbf{Initialization:} The system generates a random partition $\Pi_{\textit{ini}}$; Set the minimum transmission rate requirement $R_{min}$ ; Set the current partition $\Pi_{\textit{cur}}=\Pi_{\textit{ini}}$; Set the temporary partition $\Pi_{tem}=\Pi_{\textit{ini}}$; Set the final Nash-stable partition $\Pi_{fin}=\Pi_{\textit{ini}}$;\\
\textbf{Iteration:}
\begin{algorithmic}[1]
\WHILE {the partition converges are not achieving a Nash-stable partition}
\STATE Randomly choose a link $l\in\mathcal{L}$, and denote its current coalition as $S_p\in\Pi_{\textit{cur}}$;
\STATE Randomly choose another coalition $S_q\in(\Pi_{\textit{cur}}\cup\{\emptyset\})$, $S_p\neq S_q$; 
\IF{$\forall l^\prime, l^{\prime\prime}\in S_q$, $\exists  r_l=s_{l^\prime}$ or $s_l=r_{l^{\prime\prime}}$}
\STATE Denote the sub-channel which the links in coalition $S_q$ occupy by $c$;
\STATE Calculate the transmission rates of link $l$ and link $l^{\prime\prime}$, and denote them by $R_l^c$ and $R_{l^{\prime\prime}}^c$;
\ELSE
\STATE Go to line 11;
\ENDIF
\IF{$R_l^c\geq R_{min} \text{ and } R_{l^{\prime\prime}}^c\geq R_{min}$}
\IF{the switch operation from $S_p$ to $S_q$ satisfying $S_q\succ_l S_p$}
\STATE $\Pi_{\textit{cur}}=(\Pi_{\textit{cur}}\backslash\{S_p, S_q\})\cup\{S_p\backslash l\}\cup\{S_q\cup l\}$;
\ELSE
\STATE $\Pi_{tem}=(\Pi_{\textit{cur}}\backslash\{S_p, S_q\})\cup\{S_p\backslash l\}\cup\{S_q\cup l\}$;
\STATE Randomly choose a link $l_1\in\mathcal{L}$, and denote its current coalition as $S_{p^\prime}\in\Pi_{\textit{tem}}$;
\STATE Randomly choose another coalition $S_{q^\prime}\in(\Pi_{\textit{tem}}\cup\{\emptyset\})$, $S_{p^\prime}\neq S_{q^\prime}$; 
\IF{$\forall l_1^\prime, l_1^{\prime\prime}\in S_{q^\prime}$, $\exists r_{l_1}=s_{l_1^\prime}$ or $s_{l_1}=r_{l_1^{\prime\prime}}$}
\STATE Denote the sub-channel which the links in coalition $S_{q^\prime}$ occupy by $c^\prime$;
\STATE Calculate the transmission rates of link $l_1$ and link $l_1^{\prime\prime}$, and denote them by $R_{l_1}^{c^\prime}$ and $R_{l_1^{\prime\prime}}^{c^\prime}$;
\ELSE
\STATE Go to line 24;
\ENDIF
\IF{$R_{l_1}^{c^\prime}\geq R_{min} \text{ and } R_{l_1^{\prime\prime}}^{c^\prime}\geq R_{min}$}
\STATE $\Pi_{\textit{tem}}^\prime=(\Pi_{\textit{tem}}\backslash\{S_{p^\prime}, S_{q^\prime}\})\cup\{S_{p^\prime}\backslash l_1\}\cup\{S_{q^\prime}\cup l_1\}$;
\IF{$R(\Pi_{\textit{tem}}^\prime)>R(\Pi_{\textit{cur}})$}
\STATE $\Pi_{\textit{cur}}=\Pi_{\textit{tem}}^\prime$;
\ENDIF
\ELSE
\STATE $S_{q_1^\prime} = S_{q^\prime}$;
\STATE Randomly choose another coalition \\$S_{q^\prime}\in(\Pi_{\textit{tem}}\cup\{\emptyset\})$, $S_{q^\prime}\neq S_{q_1^\prime}$, $S_{q^\prime}\neq S_{p^\prime}$;
\STATE Go back to line 17;
\ENDIF
\ENDIF
\ELSE
\STATE $S_{q_1} = S_q$;
\STATE Randomly choose another coalition $S_q\in(\Pi_{\textit{cur}}\cup\{\emptyset\})$, $S_q\neq S_{q_1}$, $S_q\neq S_p$;
\STATE Go back to line 4;
\ENDIF
\ENDWHILE
\STATE Output $\Pi_{\textit{fin}}=\Pi_{\textit{cur}}$;
\end{algorithmic}
\end{algorithm}

The condition of finishing the loop is the partition converges can achieve a Nash-Stable partition. We prove the convergence of Algorithm~\ref{alg: Coalition Formation Algorithm} as follows
\begin{proposition}
Any initial partition $\Pi_{\textit{ini}}$, after finite switch operations, the coalition formation algorithm for sub-channel allocation will always converge to a final network partition $\Pi_{\textit{fin}}$ composed of a number of disjoint coalitions.
\end{proposition}

Due to the preference relation defined in (\ref{equation:quantify}), we can find that a single switch operation of any player $l\in\mathcal{L}$ may lead to yield an unvisited partition or a previously visited partition with a non-cooperative player $l$ (player $l$ is a singleton coalition of the new partition). If there is a non-cooperative player $l$ in partition, it need decide to join a new coalition or remain non-cooperative. If player $l$ remains non-cooperative, the current partition cannot be changed to any visited partitions in the next turn. If player $l$ decides to join a new coalition, the switch operation made by player $l$ will form an unvisited partition without any non-cooperative player. No matter how to do it, an unvisited partition will be formed. Besides, as the well known fact is the number of partitions of a set is given by the Bell number \cite{Leveraging5, Games2}, so there are finite different partitions in total.

In the case where every single switch operation can yield an unvisited partition, and the partitions consisting of the fixed players is finite, the coalition formation of the proposed algorithm will converge to a final network partition $\Pi_{\textit{fin}}$  composed of a number of disjoint coalitions after finite turns. So far the proof has been completed.

\begin{proposition}
The final partition $\Pi_{\textit{fin}}$ in our coalition formation algorithm is Nash-stable.
\end{proposition}

If the final partition $\Pi_{\textit{fin}}$ of the proposed algorithm is not Nash-stable. Consequently, there must exist a player $l\in S_p$ ($S_p\subseteq\Pi_{\textit{fin}}$) and another coalition $S_c\in\Pi_{\textit{fin}}$ such that $S_c\succ_l S_p$. Based on our proposed algorithm, player $l$ will perform a switch operation from $S_p$ to $S_c$ and form a new partition, which is contradictory to the fact that $\Pi_{\textit{fin}}$ is the final partition. Thus, the proposition that any final partition $\Pi_{\textit{fin}}$ resulting from the proposed algorithm is Nash-stable has been proved.

\section{Performance Evaluation}\label{S6}
In this section, we evaluate the performances of our proposed cooperation approach for sub-channel allocation under various patterns and system conditions. In a typical FD scenario of small cells densely deployed underlying the mmWave network, the proposed approach is compared with other approaches in order to confirm the performances.

\subsection{Simulation Setup}\label{S6-1}
In the simulations, all the BSs and user equipments are uniformly distributed in a $100m\times100m$ square area. They have the same transmission power $P_t$, the transmitters and receivers of links are randomly selected. There are 3 small cells, access links and D2D links are generated randomly, and the maximum distance of D2D links is $5m$. The minimum transmission rate $R_{min}$ for all the links are distributed differently according to different situations. The SI cancellation parameters $\beta$ for equipments are uniformly distributed in a certain range. Other parameters are shown in Table \ref{tab:parameters}.
\begin{table}[htbp]
\captionsetup{font={small}}
 \caption{\label{tab:parameters} Simulation Parameters}
 \centering
 \begin{tabular}{lcl}
  \toprule
  Parameter&Symbol&Value\\
  \midrule
  Transmission power&$P_t$&30 dBm\\
  Transceiver efficiency factor&$\eta$&0.5\\
  Path loss exponent&$n$&2\\
  Sub-channel bandwidth&$W$& 540\,MHz\\
  Background noise&$N_0$&-134dBm/MHz\\
  Maximum distance of D2D link&$d$&5m\\
  Minimum transmission rate&$R_{min}$&400 Mbit/s\\
  SI cancellation level&$\beta$&$0.5\scriptsize{\sim}1.5$\\
  Half-power beamwidth&$\theta_{-3dB}$&$30^\circ$\\
  \bottomrule
 \end{tabular}
\end{table}

\begin{figure}[t]
\centering
\subfigure[System throughput under different number of D2D links.]{
\label{fig:rate-link}
\includegraphics[width=0.45\textwidth]{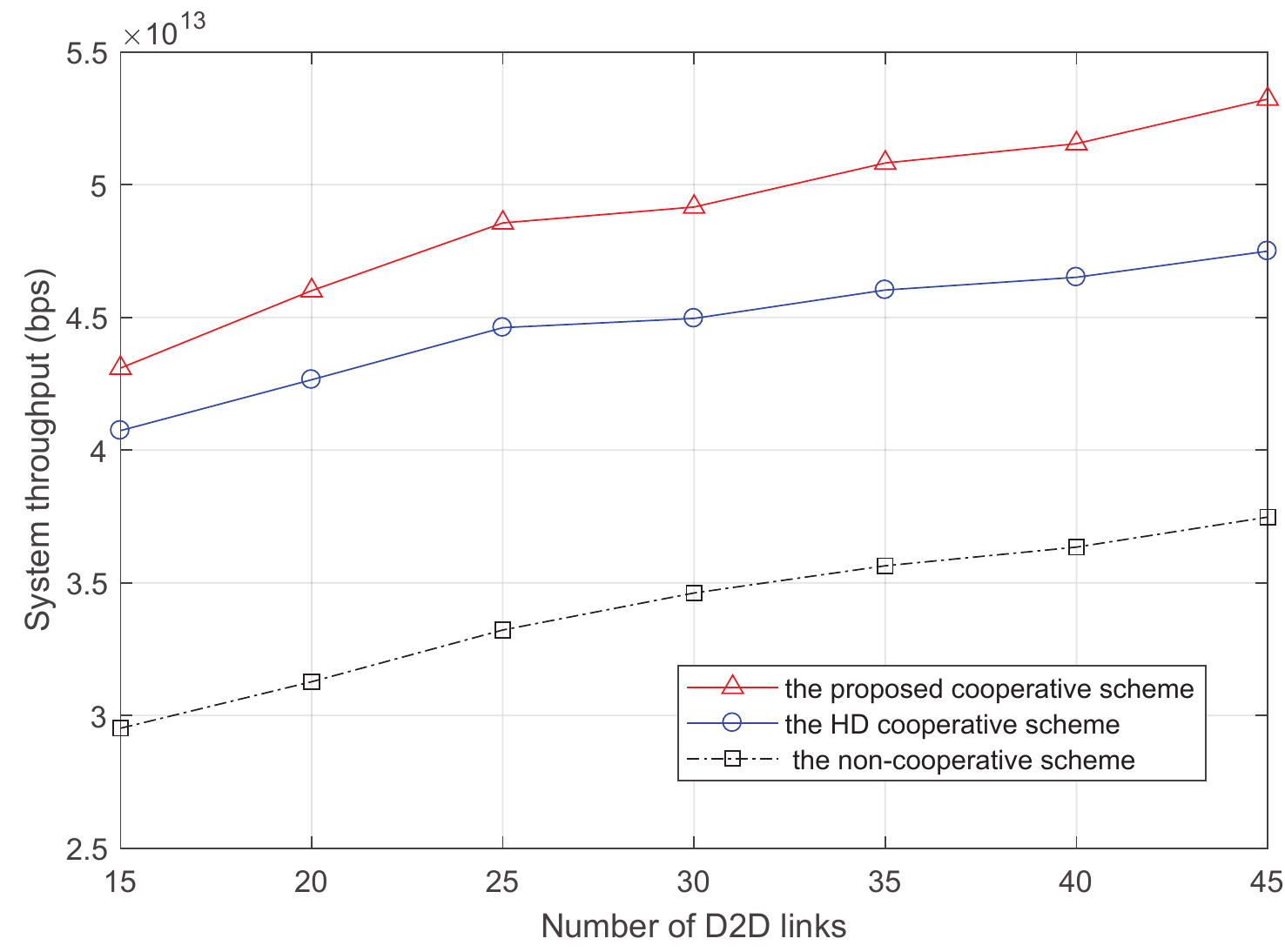}}
\subfigure[System throughput under different number of access links.]{
\label{fig:rate-alink}
\includegraphics[width=0.45\textwidth]{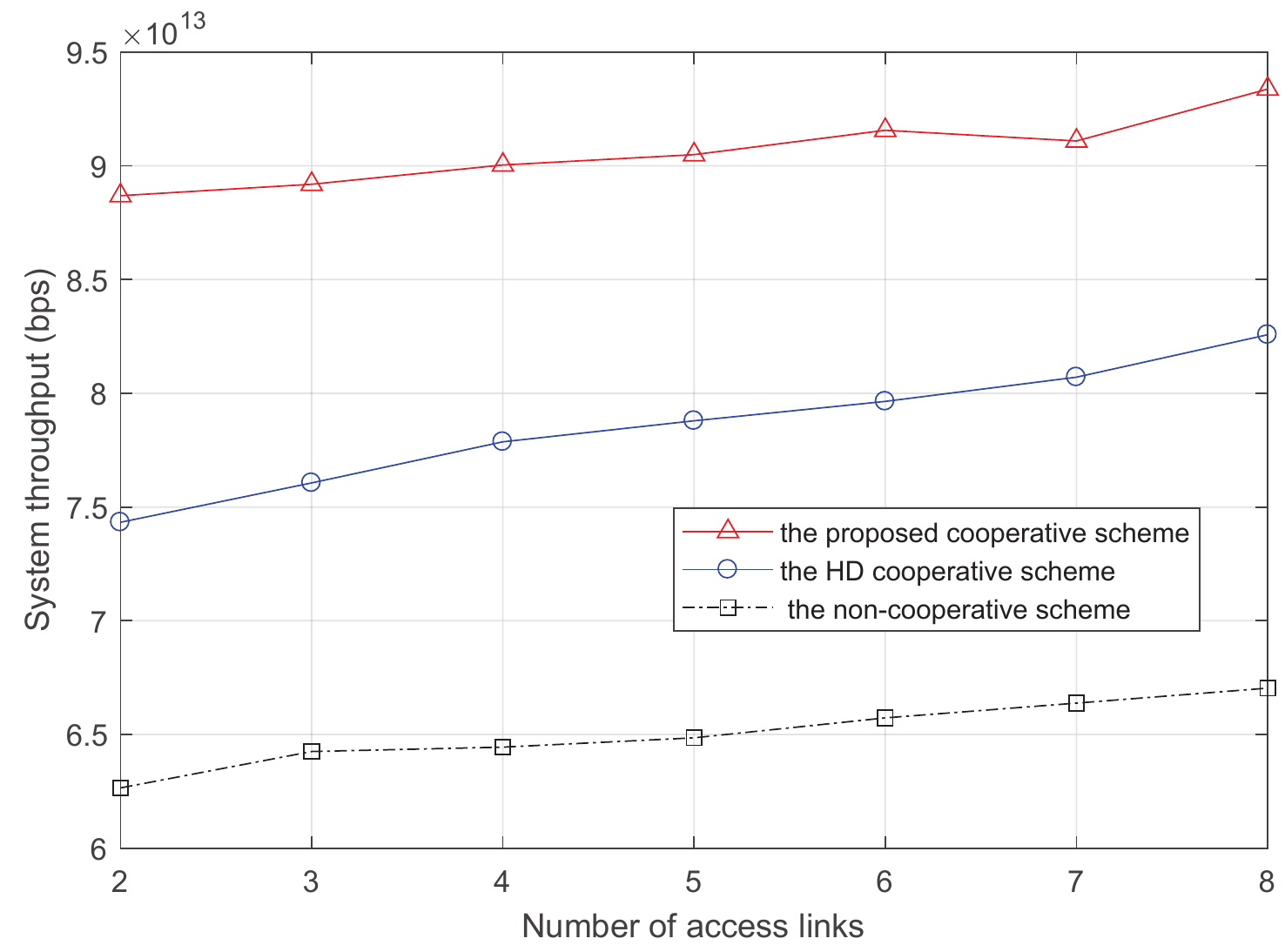}}
\captionsetup{font={small}}
\caption{System throughput under different number of D2D links and access links.}
\label{fig:System sum rate links}
\end{figure}

In order to show the advantages of our approach in terms of simulation results, we evaluate the following three performance metrics.

1)\textbf{System throughput}:  The achieved throughput of the links scheduled at the same time. The rate of link $l$ is denoted by  $R_l$, then the \emph{system throughput} $T$ can be calculated as
\begin{equation}\label{eqratio}
T=\sum_{l\in\mathcal{L}}R_l.
\end{equation}

The system throughput is in the unit of bps, and the utility of the channel allocation problem is proportional to it.

2)\textbf{Fairness}: Fairness performance is denoted by the Jain¡¯s fairness measure. The Jain¡¯s fairness measure \cite{45} can be used to determine whether the resources are fairly distributed among the links. The transmission rate of link $l$ $(l\in\mathcal{L})$ is denoted by $R_l$, then the Jain¡¯s fairness measure can be obtained as
\begin{equation}
\begin{split}
J(R_1,R_2,...,R_{|\mathcal{L}|})=\frac{(\sum_{l\in\mathcal{L}}R_l)^2}{|\mathcal{L}|\cdot\sum_{l\in\mathcal{L}}R_l^2}.\label{equation:Jain¡¯s fairness}
\end{split}
\end{equation}

3)\textbf{Number of switch operations}: Average number of the switch operations when the partition is a Nash-stable partition in proposed algorithm.

In order to show the advantages of the proposed coalition game algorithm ($\textbf{the proposed cooperative scheme}$) in the system, we compare it with following three schemes.

1)\textbf{the HD cooperative scheme}: It uses the same coalition game algorithm with the proposed cooperative scheme to do the sub-channel allocation, but it only allows HD communication with HD requirements.

2)\textbf{the non-cooperative scheme}: Random Allocation, where the sub-channels are allocated to each link randomly.

3)\textbf{optimal scheme}: Optimal Solution, which achieves an optimal sub-channel allocation scheme with the exhaustive search. The optimal scheme complexity is NP-hard, and optimal solution only can be found with few links in a small network.

To be more reliable, the simulation results are obtained by 200 independent experiments.

\subsection{Comparison with Existing Schemes} \label{S6-2}

\subsubsection{System Throughput}
In Fig.\ref{fig:rate-link} and Fig.\ref{fig:rate-alink}, we plot the system throughput under different number of D2D links and different number of access links, respectively. There are 5 access links in Fig.\ref{fig:rate-link} and 30 D2D links in Fig.\ref{fig:rate-alink}. The number of sub-channels is set to 5 in two simulation diagrams. With the increase in the links, the system throughput of three schemes all increase. The proposed cooperative scheme always shows superior performance compared with other schemes. Compared with the HD cooperative scheme, the proposed FD scheme can simultaneously transmit more links in the same sub-channel to get the larger system throughput. For the non-cooperative scheme, its system throughput is much worse than proposed FD scheme. Because it is so difficult for non-cooperative scheme to obtain the best way of sub-channels allocation by one time sub-channel allocation. Compared with Fig.\ref{fig:rate-link}, the growth trend of Fig.\ref{fig:rate-alink} is slower. This is because the access links of the abscissa in Fig.\ref{fig:rate-alink} increases more slowly than the D2D links of the abscissa in Fig.\ref{fig:rate-link}. Less increase in the number of links each time leads to slower system throughput growth.

\begin{figure}[t]
  \begin{center}
  \includegraphics[width=8cm]{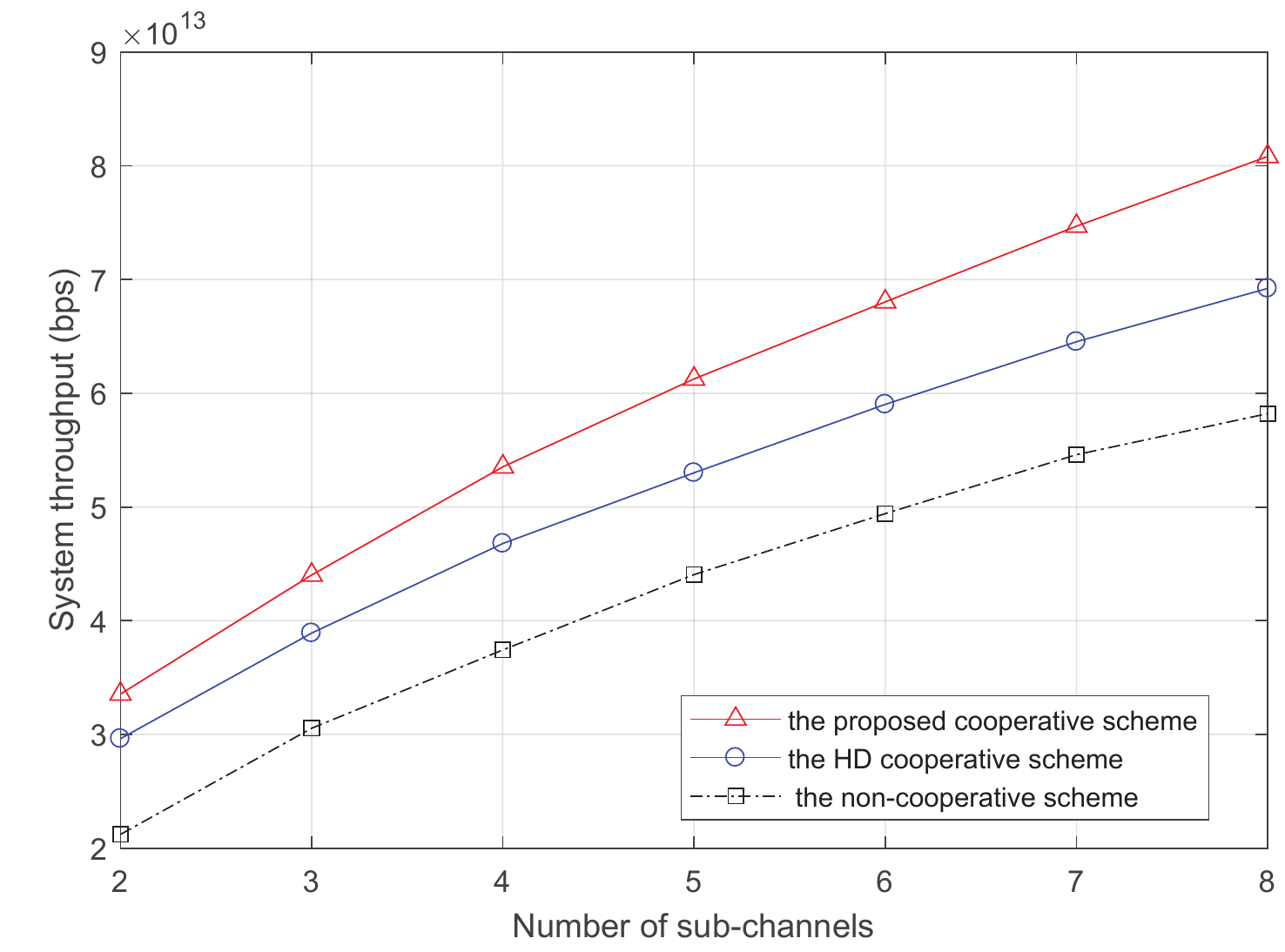}
  \captionsetup{font={small}}
  \caption{System throughput under different number of sub-channels.} \label{fig:rate-channel}
  \end{center}
\end{figure}
\begin{figure}[t]
  \begin{center}
  \includegraphics[width=8cm]{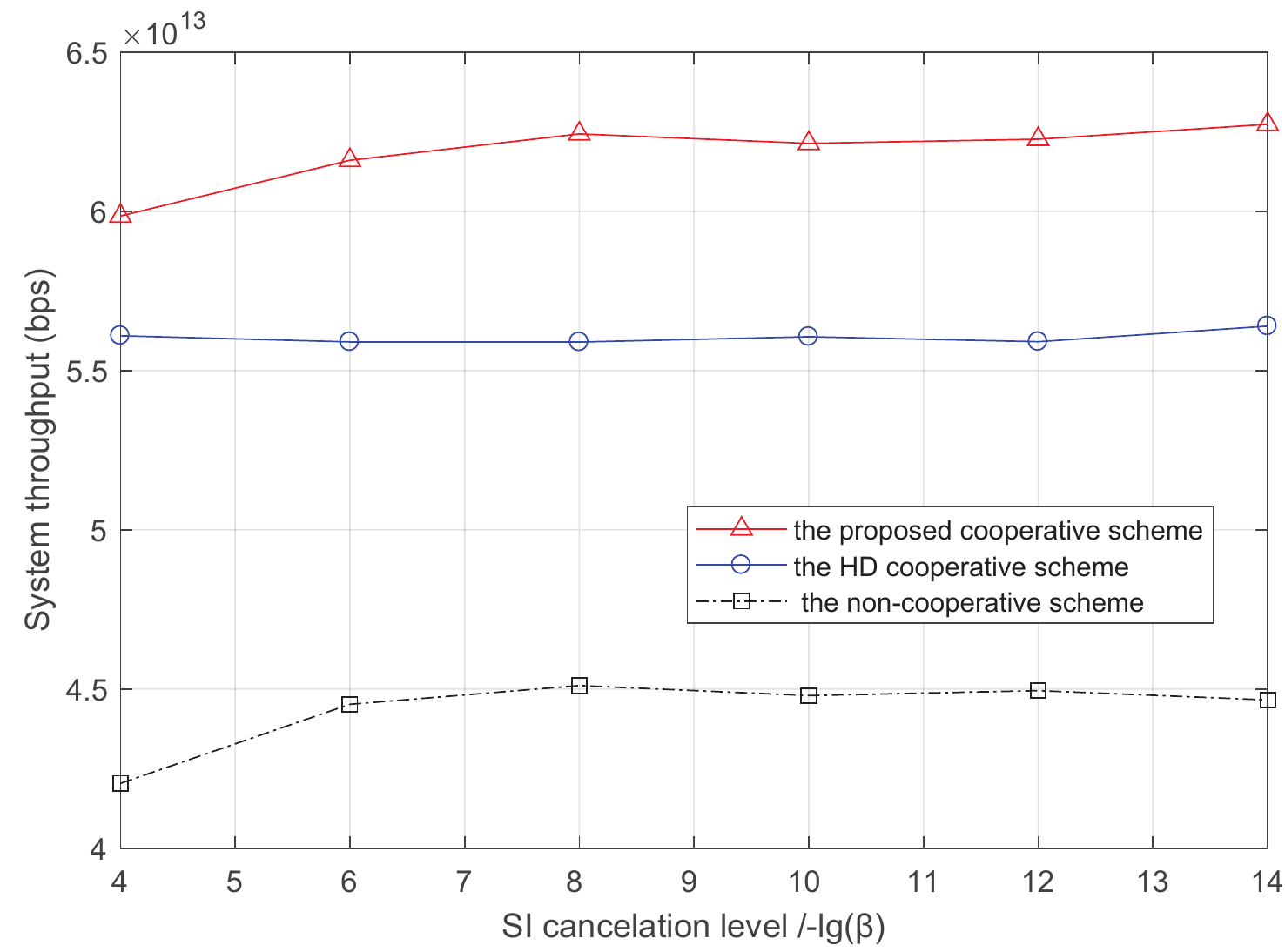}
  \captionsetup{font={small}}
  \caption{System throughput under different SI cancellation levels.} \label{fig:rate-h}
  \end{center}
\end{figure}

In Fig.\ref{fig:rate-channel}, we plot the system throughput under different number of sub-channels. There are 3 access links and 70 D2D links. We can observe that the proposed cooperative scheme also has a superior performance in terms of the system throughput among three schemes. With the increasing of sub-channel, the system throughput of all different schemes increase stably. When the number of sub-channels becomes larger, there will be less links in each sub-channel and less interference between the links, then the transmission rate of each link can be larger. When the number of sub-channels is 8, the proposed FD scheme improves the system throughput by about 15.9\% compared with the HD scheme, and by about 37.9\% compared with the non-cooperative allocation scheme.

In Fig.\ref{fig:rate-h}, we simulate the the system throughput under different magnitudes of SI cancellation levels. The number of D2D links is 15, the number of access links is 5, and the number of sub-channels is set to 5. The abscissa $x$ is the magnitude of $\beta$. For example, when $x = 8$, $\beta$ is uniformly distributed in $ 0.5 \times 10^{-8} \scriptsize{\sim} 1.5 \times 10^{-8}$. We can find that the performance of the proposed cooperative scheme is better than other schemes. The system throughput of the HD scheme has no change under different SI cancellation levels, because HD algorithm does not need SI cancellation. For other two FD schemes, the cooperative scheme and the non-cooperative scheme, the SI cancelation level has significant impacts on their performances. When $x$ is small, that is, the SI cancelation level is low, the performances of cooperative scheme and non-cooperative scheme are not ideal. As $x$ becomes larger, the RSI of the transmission is less and the RSI has a negative impact on the transmission rate, so the system throughput of two FD schemes increase. When SI cancellation level reaches $10^{-8}$ magnitude, the system throughput of FD schemes are nearly unchanged with $x$ increasing. At this moment, the RSI of the transmission is less enough and its effect on rate is very small. In other simulations, we set $\beta$ to $0.5\times10^{-8}\scriptsize{\sim}1.5\times10^{-8}$, which can get good performances of schemes and achieve normal transmission.

\subsubsection{Fairness Performance}
\begin{figure}[t]
\centering
\subfigure[Fairness under different number of D2D links.]{
\label{fig:fair-link}
\includegraphics[width=0.45\textwidth]{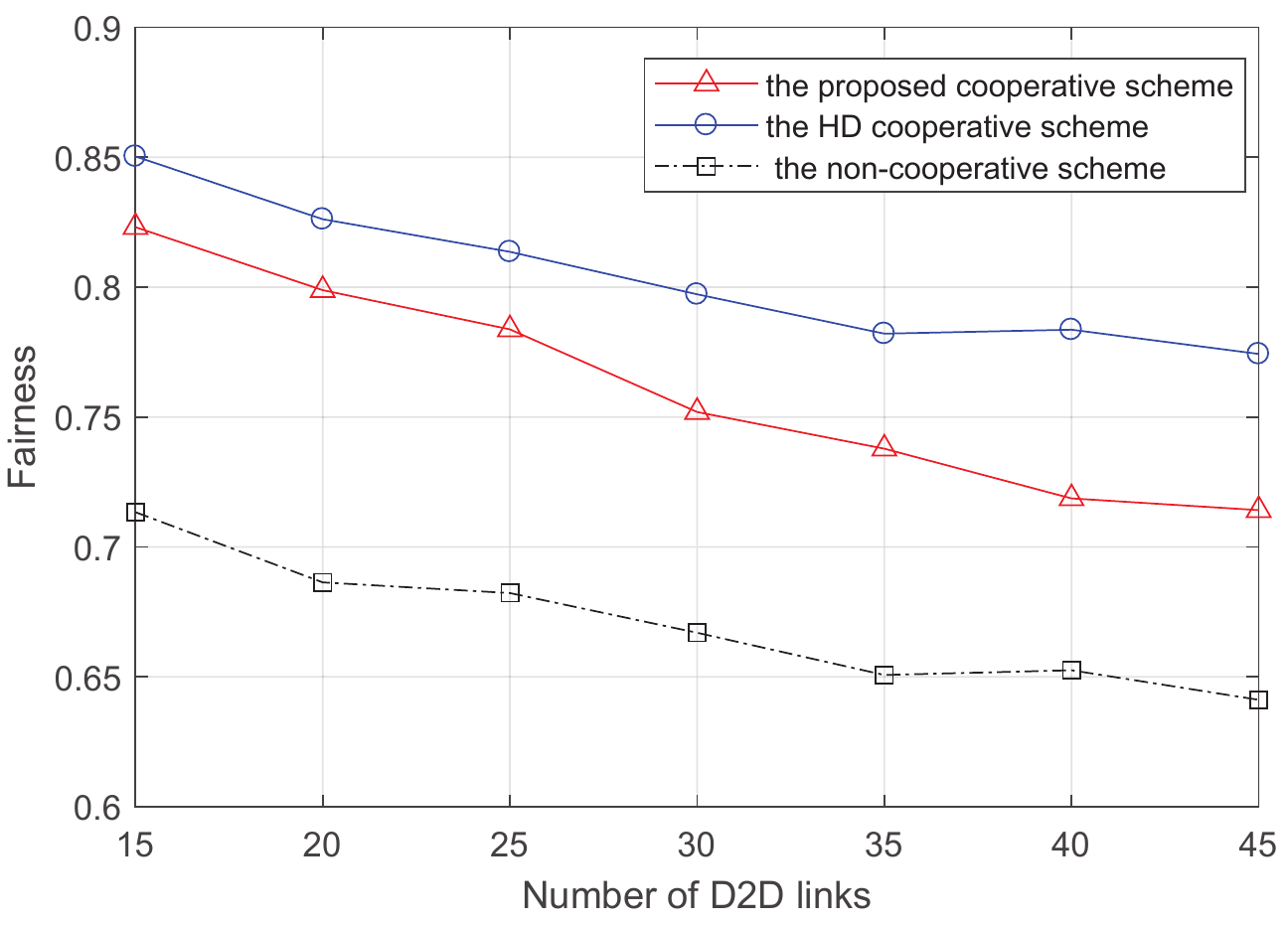}}
\subfigure[Fairness under different number of access links.]{
\label{fig:fair-alink}
\includegraphics[width=0.45\textwidth]{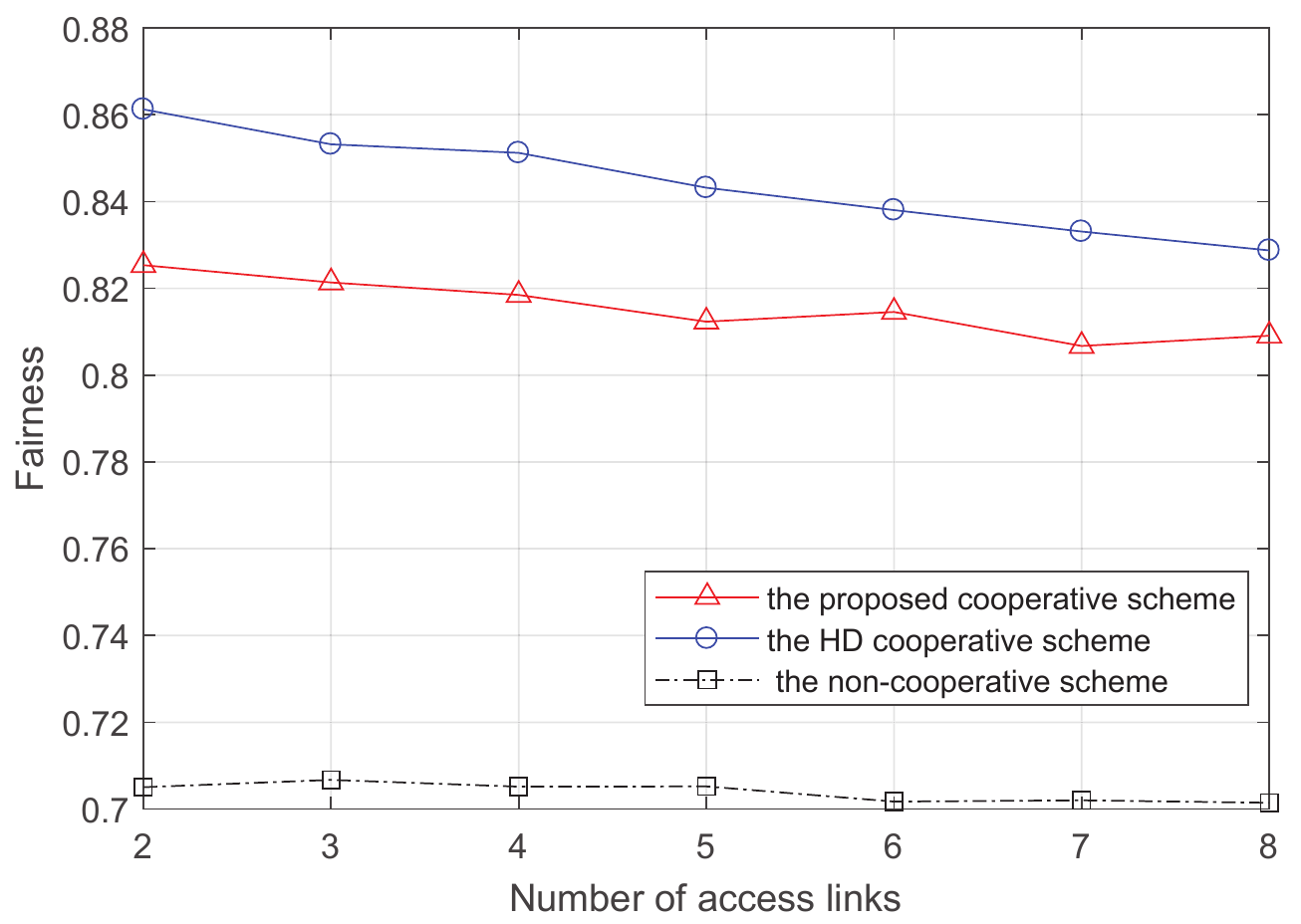}}
\captionsetup{font={small}}
\caption{Fairness of different allocation algorithms under different number of D2D links and access links.}
\label{fig:Fairness links}
\end{figure}

In Fig.\ref{fig:fair-link} and Fig.\ref{fig:fair-alink}, we plot the fairness performances of different sub-channel allocation schemes under different number of D2D links and different number of access links respectively. The number of access links is set to 5 in Fig.\ref{fig:fair-link}, the number of D2D links is set to 30 in Fig.\ref{fig:fair-alink} and the number of sub-channels is set to 5 in two diagrams. We can observe that our scheme has the better fairness performance compared with the non-cooperative scheme, but it is not better than the HD cooperative scheme. Nevertheless, we can not get a conclusion that the proposed cooperative scheme has poor performance on system fairness performance compared with HD scheme. In fact, the links simultaneously scheduled by HD algorithm are less than these by FD scheme, and less links simultaneously transmitting has advantage in the fairness. With the increasing of the links, the fairness indexes of all schemes decrease with more interferences between links.

\begin{figure}[t]
  \begin{center}
  \includegraphics[width=8cm]{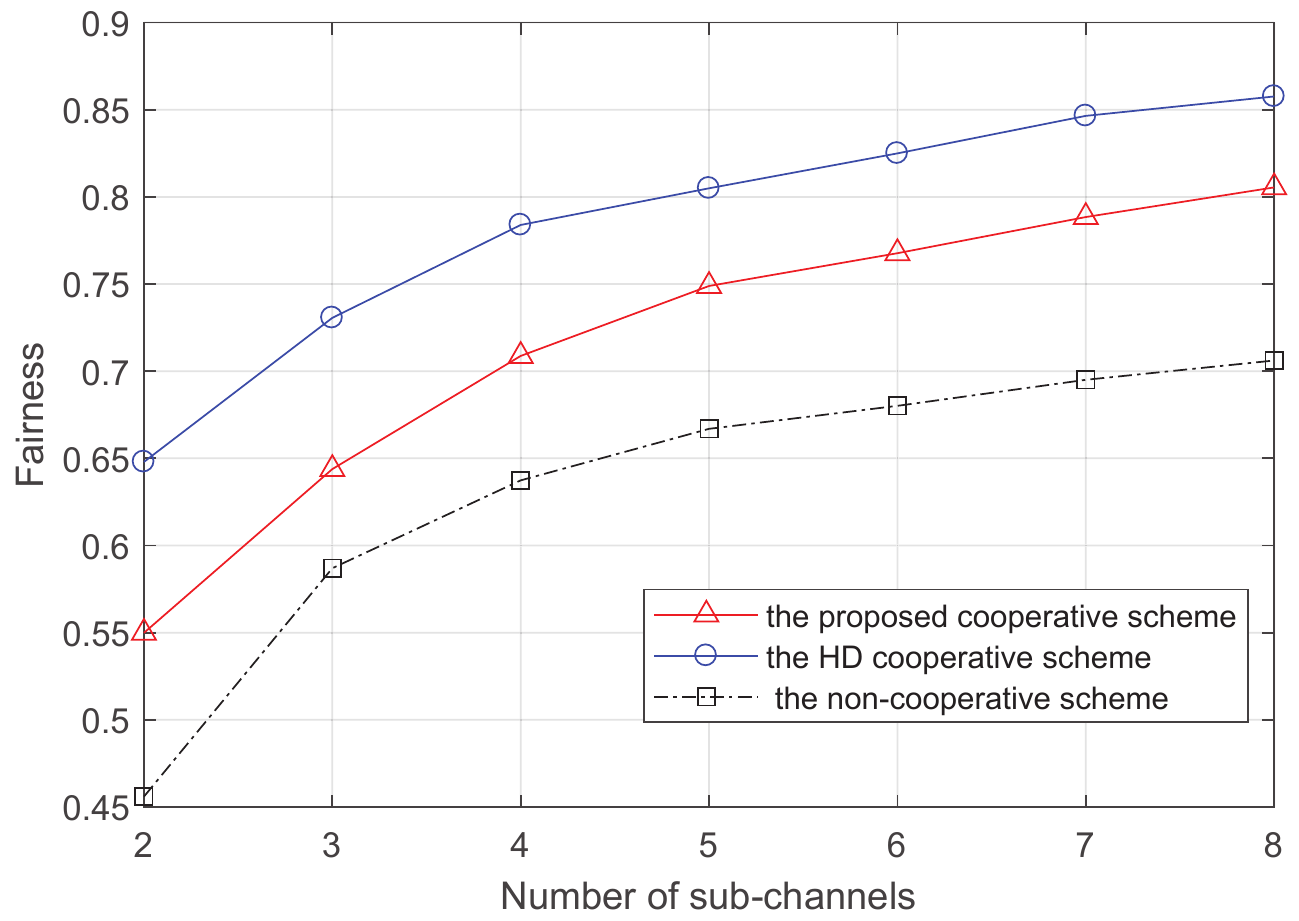}
  \captionsetup{font={small}}
  \caption{Fairness of different allocation algorithms under different number of sub-channels.} \label{fig:fair-channel}
  \end{center}
\end{figure}

In Fig.\ref{fig:fair-channel}, we plot the fairness performance of different sub-channels allocation algorithms under different numbers of sub-channels. There are 3 access links and 70 D2D links. We can also find that our proposed scheme has a better fairness performance compared with the non-cooperative scheme, but it is not better than HD scheme. The reasons are similar to Fig.\ref{fig:fair-link}. With the increasing of the number of sub-channel, the fairness indexes of all schemes increase. The sub-channel is more, the sub-channels allocation is fairer.

\subsubsection{Impact of the minimum transmission rate}

$R_{min}$ is an important parameter of the fairness and has a significant impact on the proposed cooperative scheme. Next, we study the impacts of $R_{min}$ on the proposed cooperative scheme.

\begin{figure}[t]
  \begin{center}
  \includegraphics[width=8cm,height=5.8cm]{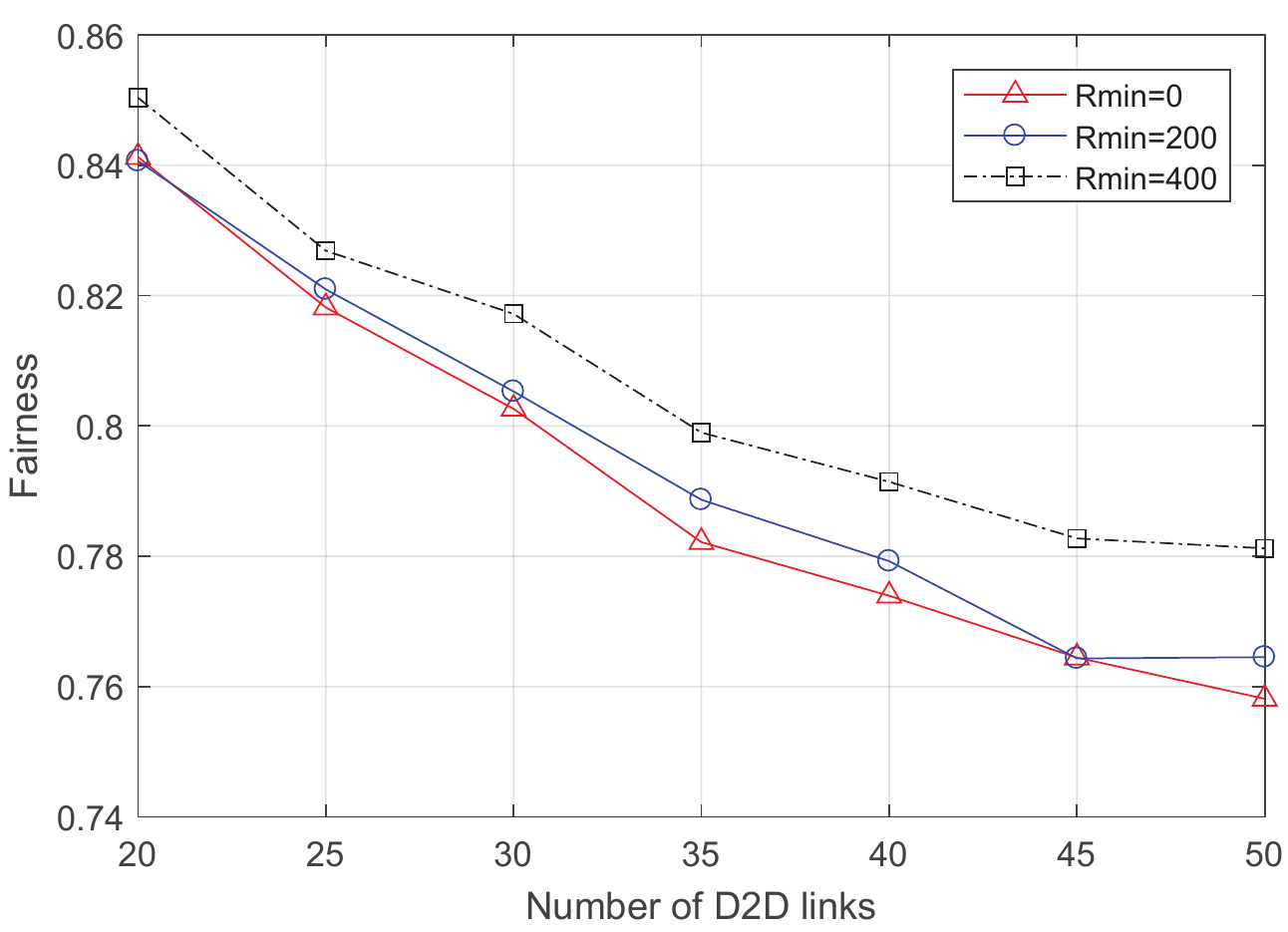}
  \captionsetup{font={small}}
  \caption{Fairness under different number of D2D links.} \label{fig:fair-r}
  \end{center}
\end{figure}

In Fig.\ref{fig:fair-r}, we plot the fairness performance of the proposed cooperative allocation scheme under different number of D2D links, $R_{min}$ equals to 0 Mbit/s, 200 Mbit/s, and 400 Mbit/s, respectively. The number of access links is 5 and the number of sub-channels is 5. We can observe that the proposed cooperative scheme can become fairer and fairer with the growth of $R_{min}$. When the value of $R_{min}$ is large, the links with small rates are less and the unfairness caused by these links can be alleviated. Thus, the fairness becomes better with $R_{min}$ becoming larger. As envisioned by the above content, $R_{min}$ is beneficial to the fairness of the links.

\begin{figure}[t]
  \begin{center}
  \includegraphics[width=8cm,height=7.1cm]{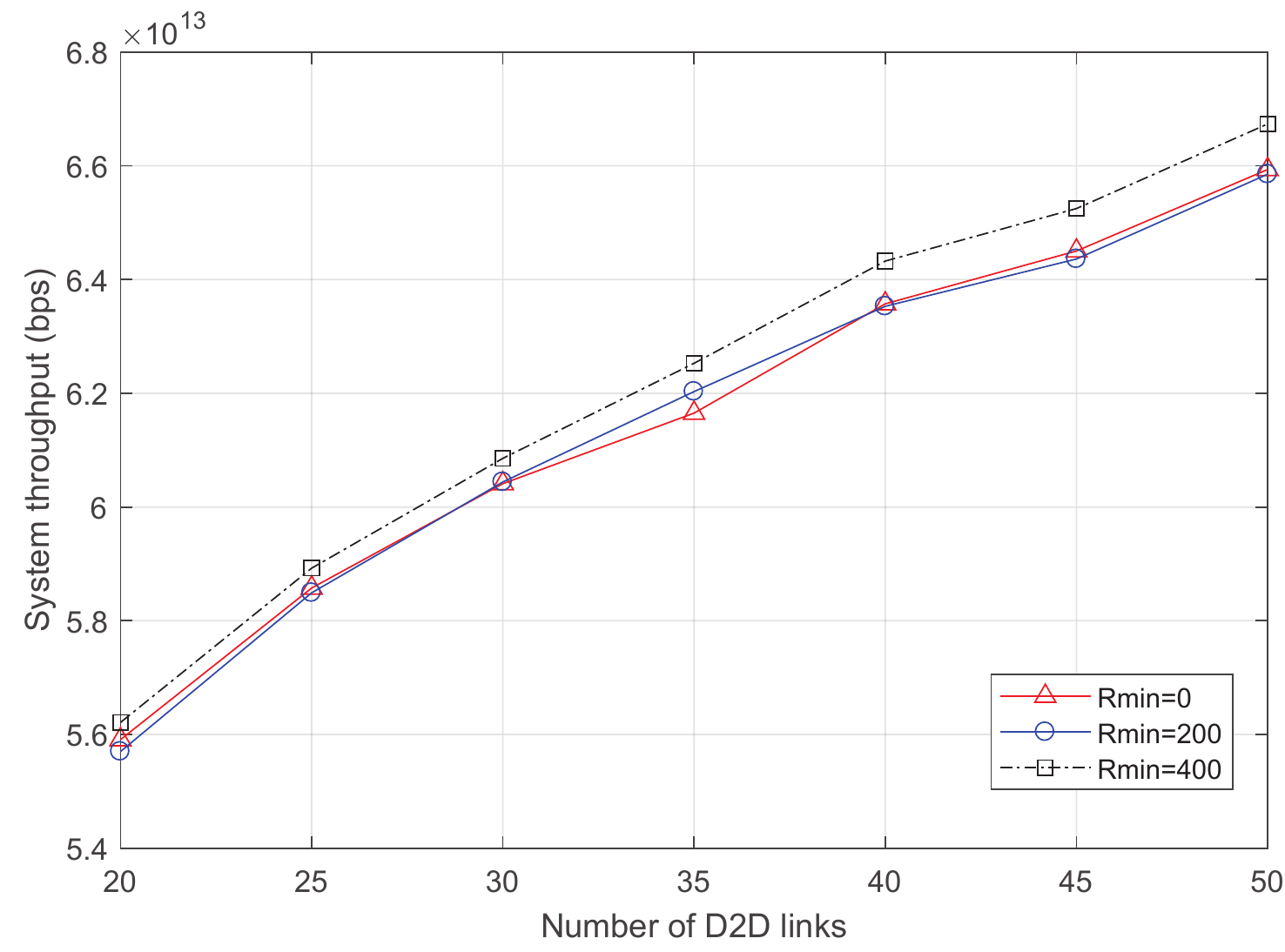}
  \captionsetup{font={small}}
  \caption{System throughput under different number of D2D links} \label{fig:rate-r}
  \end{center}
\end{figure}

In Fig.\ref{fig:rate-r}, we plot the system throughput of the proposed allocation scheme under different number of D2D links with different $R_{min}$, $R_{min}$ equals to 0 Mbit/s, 200 Mbit/s, and 400 Mbit/s, respectively. All parameters are same as parameters in Fig.\ref{fig:fair-r}. With the increasing of $R_{min}$, there is no regular trend of fairness changing. In fact, the effect of $R_{min}$ on the system throughput is undetermined. When the proposed cooperative scheme does sub-channels allocation, it makes decisions based on the principle of maximizing the sum throughput of the entire system, and mainly pursuing the overall benefits. As $R_{min}$ increases, the requirement on the transmission rate of each link becomes higher, the system throughput can be bigger or smaller uncertainly.


Therefore, we can get a conclusion from Fig.\ref{fig:rate-r} and Fig.\ref{fig:fair-r}, the value of $R_{min}$ should be appropriate. If $R_{min}$ is small, the performance of fairness is not good enough. On the contrary, if the value of $R_{min}$ is big, the coalition formation algorithm may not find a suitable partition of all the links. In other simulations, we set $R_{min}$ to 400 Mbit/s, which does not affect the performances of the algorithms.

\subsection{Comparison with Optimal Solution} \label{S6-3}
\begin{figure}[t]
  \begin{center}
  \includegraphics[width=8cm,height=7cm]{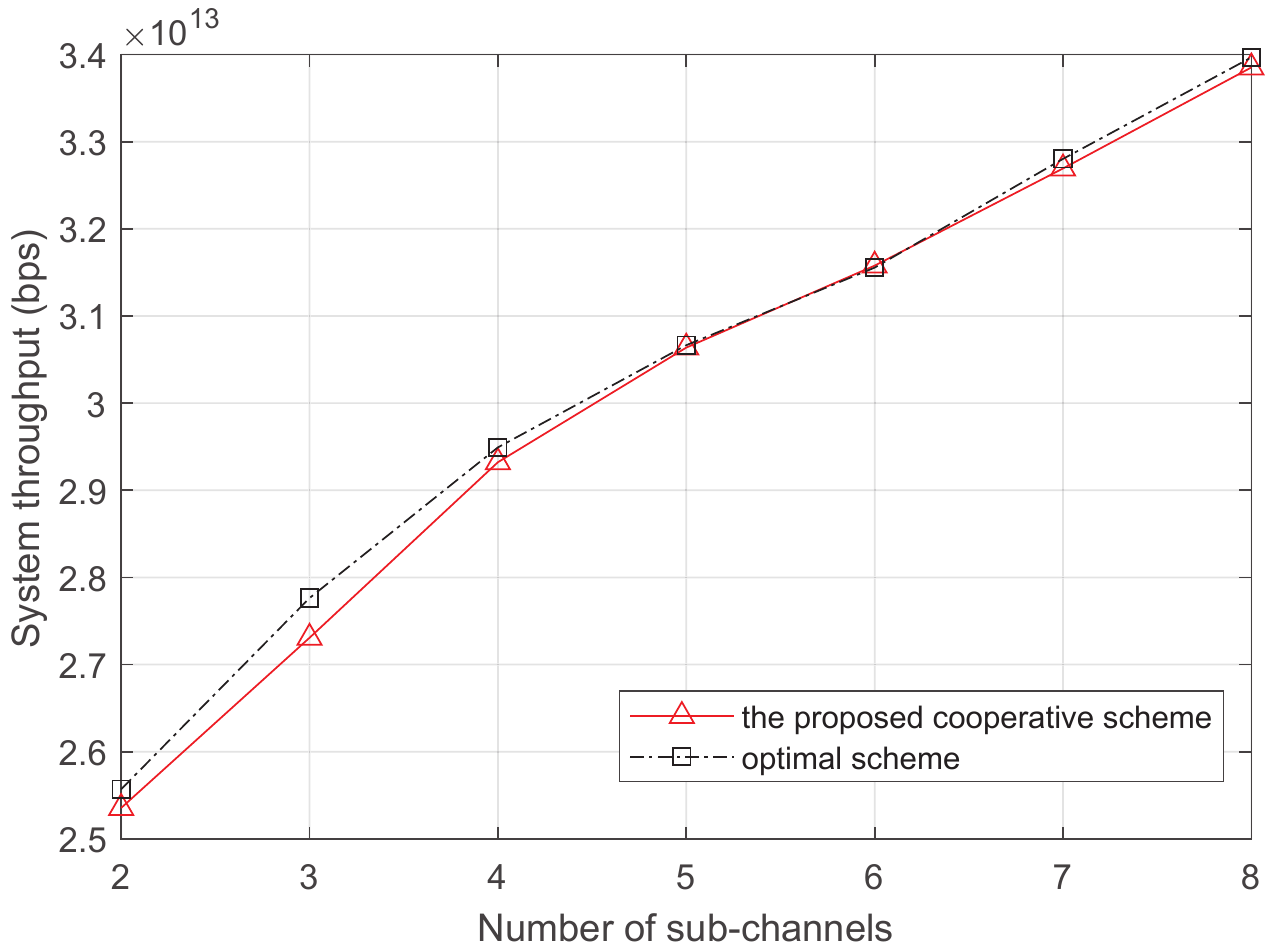}
  \captionsetup{font={small}}
  \caption{System throughput under different number of D2D links.} \label{fig:optimal solution}
  \end{center}
\end{figure}
\begin{figure}[t]
  \begin{center}
  \includegraphics[width=8cm,height=7.1cm]{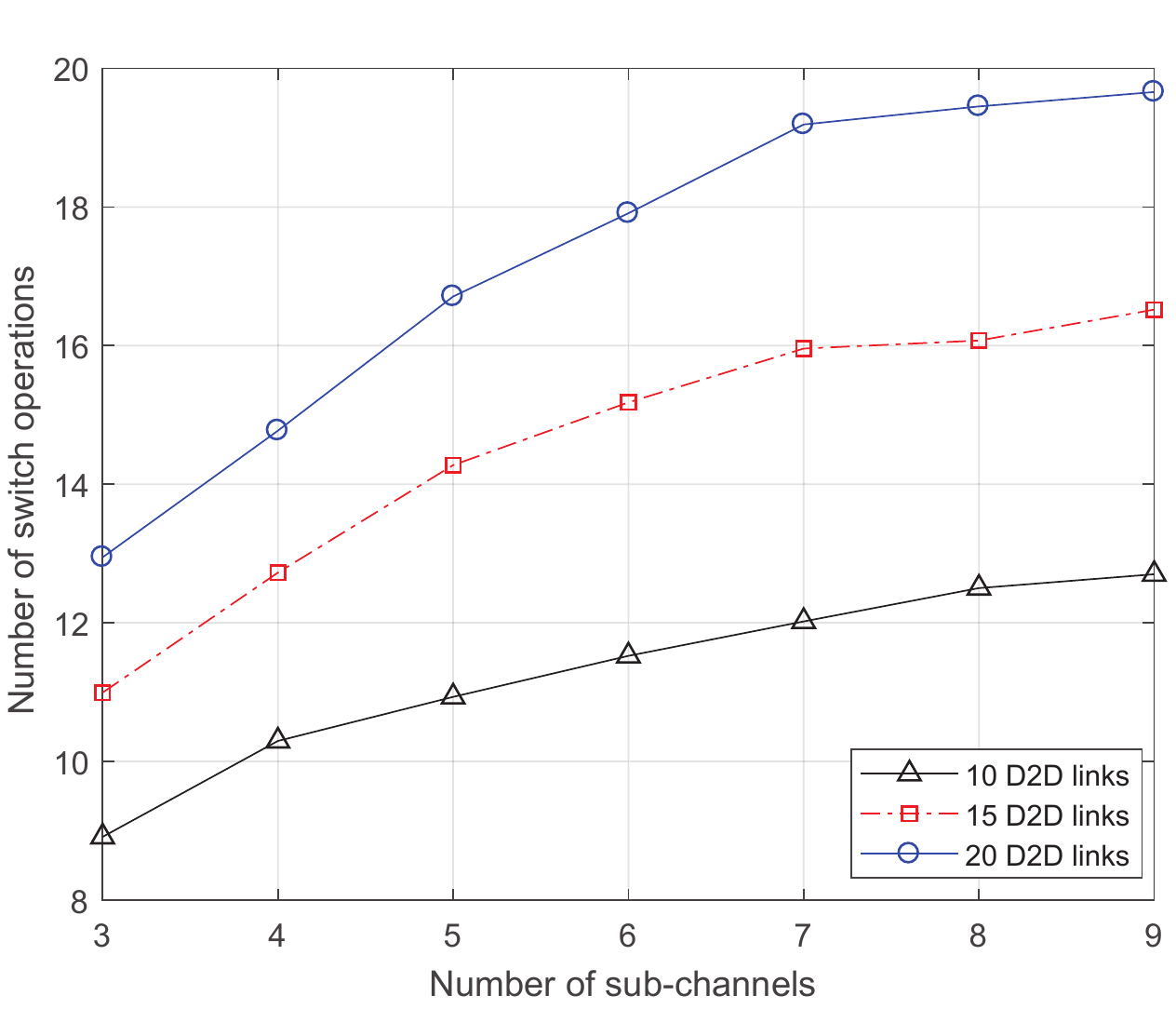}
  \captionsetup{font={small}}
  \caption{Number of switch operations under different number of sub-channels.} \label{fig:switch}
  \end{center}
\end{figure}

To demonstrate our proposed scheme can achieve near-optimal performance on system throughput, we compare it with the optimal solution (exhaustive search method). Since it is NP-hard to obtain the optimal solution for general cases, we only compare them in a small scale network with several sub-channels and links. In this comparison, the number of access links is 3 and the number of sub-channels is set to 3.

In Fig.\ref{fig:optimal solution}, we plot the system throughput of the optimal solution scheme and proposed scheme under different number of D2D links. As we can see, the gap between the optimal solution and our scheme is negligible. When the number of links is 8, the gap is only about 0.6\%. The proposed scheme is able to approximate the optimal solution of the sub-channel allocation problem with a very small gap.

Compared with optimal solution, the proposed scheme not only has a superior performance on system throughput, the complexity is also low. We illustrate the complexity of the proposed algorithm as follows.

In Fig.\ref{fig:switch}, we show the system convergence rate in terms of the average numbers of switch operations under different number of sub-channels. We investigate three cases, with the number of D2D links is set to 10, 15, and 20, respectively, the number of access links is 5. We can observe that the average numbers of switch operations increase with the number of links increasing. More links need to be allocated, more switch operations are needed. Whatever the number of D2D links is, the average numbers of switch operations are limited and the small number of switch operations implies that the proposed cooperative scheme can do the sub-channel allocation with limited complexity. Consequently, the low complexities for different numbers of links imply the network scale of the proposed scheme in practical system can be much larger than the network scale we set in the simulations.


\section{Conclusion}\label{S7} 
In this paper, we focus on the problem of optimal sub-channel allocation for the links in the scenario of multiple mmWave small cells densely deployed. To maximize the system throughput, we model the problem as a coalition formation game. Then we propose a coalition formation algorithm for sub-channel allocation, and it fully exploits FD communications and concurrent transmission to improve network performances. Besides, we proposed a threshold of the minimum transmission rate to enhance the fairness of our proposed scheme. After that, we prove the convergence of our algorithm theoretically. Extensive simulations show that the throughput of the proposed scheme has an obviously superior performance compared with other schemes, and the low complexity of our scheme is demonstrated by the average number of switch operations. However, due to the directional antenna setting of device in the paper, the number of links that each user device can transmit and receive at the same time is limited. In the future work, we will consider the D2D communications with simultaneous multi-link transmitting and receiving by multiple antennas and further increase the system throughput. Besides, we will do the performance verification of our algorithm on the actual system platform to further demonstrate the practicality of our channel allocation scheme.


\bibliographystyle{IEEEtran}

\vspace*{108mm}
\begin{IEEEbiography}[{\includegraphics[width=1in,height=1.25in,clip,keepaspectratio]{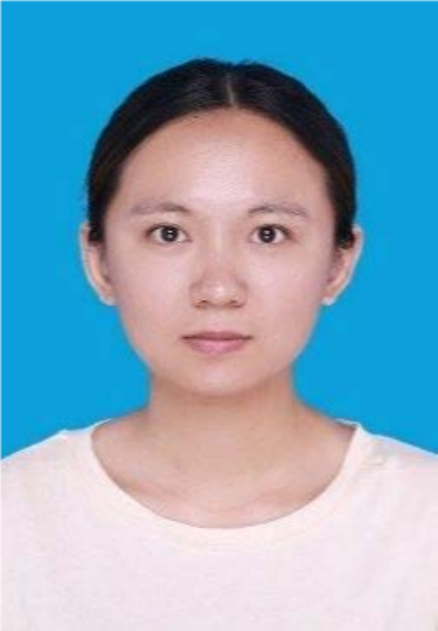}}]{Yibing Wang}
(18111034@bjtu.edu.cn) received B.E. degree in Beijing Jiaotong University, China in 2017, and she is pursuing her Ph.D. degree in Beijing Jiaotong University. Her research interests include millimeter wave communications, device-to-device communication and VANETs.
\vspace*{32mm}
\end{IEEEbiography}

\begin{IEEEbiography}[{\includegraphics[width=1in,height=1.25in,clip,keepaspectratio]{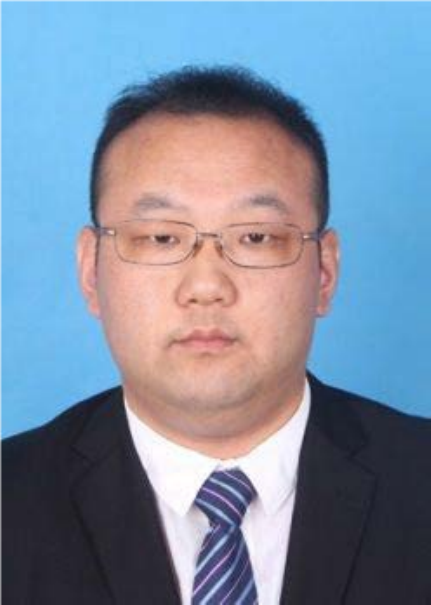}}]{Yong Niu}
(M'17) received the B.E. degree in Electrical Engineering from Beijing Jiaotong University, China, in 2011, and the Ph.D. degree in Electronic Engineering from Tsinghua University, Beijing, China, in 2016.

From 2014 to 2015, he was a Visiting Scholar with the University of Florida, Gainesville, FL, USA. He is currently an Associate Professor with the State Key Laboratory of Rail Traffic Control and Safety, Beijing Jiaotong University. His research interests are in the areas of networking and communications, including millimeter wave communications, device-to-device communication, medium access control, and software-defined networks. He received the Ph.D. National Scholarship of China in 2015, the Outstanding Ph.D. Graduates and Outstanding Doctoral Thesis of Tsinghua University in 2016, the Outstanding Ph.D. Graduates of Beijing in 2016, and the Outstanding Doctorate Dissertation Award from the Chinese Institute of Electronics in 2017. He has served as Technical Program Committee member for IWCMC 2017, VTC2018-Spring, IWCMC 2018, INFOCOM 2018, and ICC 2018. He was the Session Chair for IWCMC 2017. He was the recipient of the 2018 International Union of Radio Science Young Scientist Award.
\end{IEEEbiography}

\begin{IEEEbiography}[{\includegraphics[width=1in,height=1.25in,clip,keepaspectratio]{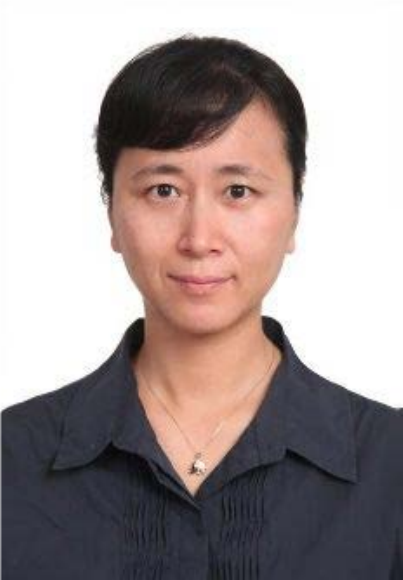}}]{Hao Wu}
(hwu@bjtu.edu.cn) received her Ph.D. degree in information and communication engineering from Harbin Institute of Technology in 2000. She is currently a full professor with the State Key Lab of Rail Traffic Control and Safety at Beijing Jiaotong University (BJTU), China. She has published more than 100 papers in international journals and conferences. Her research interests include Intelligent Transportation Systems (ITS), security and QoS issues in wireless networks (VANETs, MANETs and WSNs), wireless communications, and Internet of Things (IoT). She is a member of IEEE and a reviewer of its major conferences and journals in wireless networks and security.
\end{IEEEbiography}

\begin{IEEEbiography}[{\includegraphics[width=1in,height=1.25in,clip,keepaspectratio]{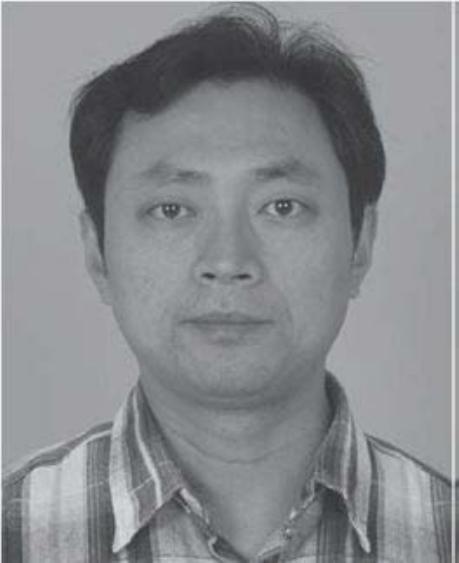}}]{Bo Ai}
received the M.S. and Ph.D. degrees from Xidian University, China. He studies as a Post-Doctoral Student at Tsinghua University. He was a Visiting Professor with the Electrical Engineering Department, Stanford University, in 2015. He is currently with Beijing Jiaotong University as a Full Professor and a Ph.D. Candidate Advisor. He is the Deputy Director of the State Key Lab of Rail Traffic Control and Safety and the Deputy Director of the International Joint Research Center. He is one of the main people responsible for the Beijing Urban Rail Operation Control System, International Science and Technology Cooperation Base. He is also a Member, of the Innovative Engineering Based jointly granted by the Chinese Ministry of Education and the State Administration of Foreign Experts Affairs. He was honored with the Excellent Postdoctoral Research Fellow by Tsinghua University in 2007.

He has authored/co-authored eight books and published over 300 academic research papers in his research area. He holds 26 invention patents. He has been the research team leader for 26 national projects. His interests include the research and applications of channel measurement and channel modeling, dedicated mobile communications for rail traffic systems. He has been notified by the Council of Canadian Academies that, based on Scopus database, he has been listed as one of the Top 1\% authors in his field all over the world. He has also been feature interviewed by the IET Electronics Letters. He has received some important scientific research prizes.

Dr. Ai is a fellow of the Institution of Engineering and Technology. He is an Editorial Committee Member of the Wireless Personal Communications journal. He has received many awards, such as the Outstanding Youth Foundation from the National Natural Science Foundation of China, the Qiushi Outstanding Youth Award by the Hong Kong Qiushi Foundation, the New Century Talents by the Chinese Ministry of Education, the Zhan Tianyou Railway Science and Technology Award by the Chinese Ministry of Railways, and the Science and Technology New Star by the Beijing Municipal Science and Technology Commission. He was a co-chair or a session/track chair for many international conferences. He is an IEEE VTS Beijing Chapter Vice Chair and an IEEE BTS Xi'an Chapter Chair. He is the IEEE VTS Distinguished Lecturer. He is an Editor of the IEEE TRANSACTIONS ON CONSUMER ELECTRONICS. He is the Lead Guest Editor of Special Issues of the IEEE TRANSACTIONS ON VEHICULAR TECHNOLOGY, the IEEE ANTENNAS AND WIRELESS PROPAGATION LETTERS, and the International Journal of Antennas and Propagation.
\end{IEEEbiography}

\begin{IEEEbiography}[{\includegraphics[width=1in,height=1.25in,clip,keepaspectratio]{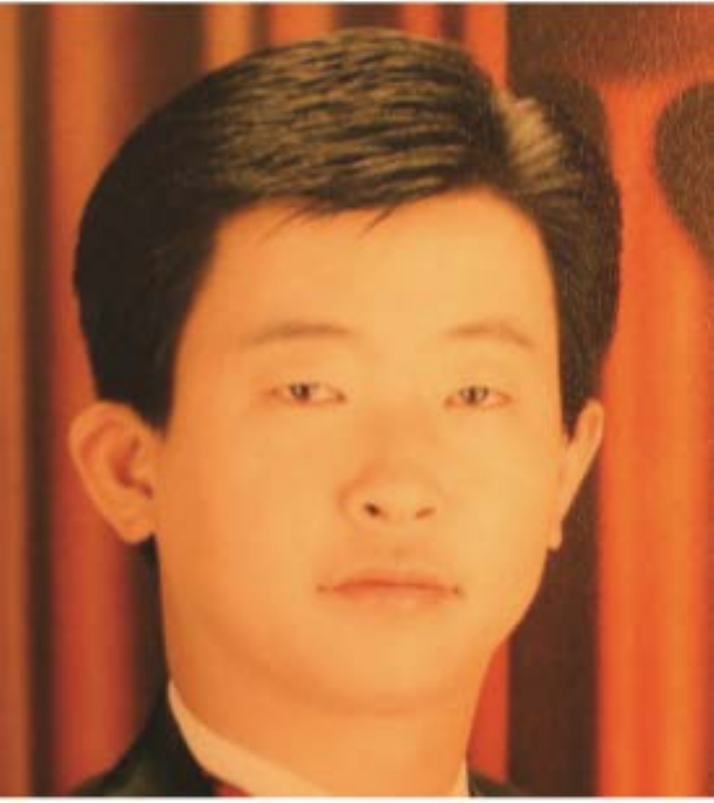}}]{Zhu Han}
(S¡¯01¨CM¡¯04-SM¡¯09-F¡¯14) received the B.S. degree in electronic engineering from Tsinghua University, in 1997, and the M.S. and Ph.D. degrees in electrical and computer engineering from the University of Maryland, College Park, in 1999 and 2003, respectively.

From 2000 to 2002, he was an R\&D Engineer of JDSU, Germantown, Maryland. From 2003 to 2006, he was a Research Associate at the University of Maryland. From 2006 to 2008, he was an assistant professor at Boise State University, Idaho. Currently, he is a John and Rebecca Moores Professor in the Electrical and Computer Engineering Department as well as in the Computer Science Department at the University of Houston, Texas. He is also a Chair professor in National Chiao Tung University, ROC. His research interests include wireless resource allocation and management, wireless communications and networking, game theory, big data analysis, security, and smart grid. Dr. Han received an NSF Career Award in 2010, the Fred W. Ellersick Prize of the IEEE Communication Society in 2011, the EURASIP Best Paper Award for the Journal on Advances in Signal Processing in 2015, IEEE Leonard G. Abraham Prize in the field of Communications Systems (best paper award in IEEE JSAC) in 2016, and several best paper awards in IEEE conferences. Currently, Dr. Han is an IEEE Communications Society Distinguished Lecturer from 2015-2018, and ACM distinguished Member since 2019. Dr. Han is 1\% highly cited researcher since 2017 according to Web of Science.
\end{IEEEbiography}

\begin{IEEEbiography}[{\includegraphics[width=1in,height=1.25in,clip,keepaspectratio]{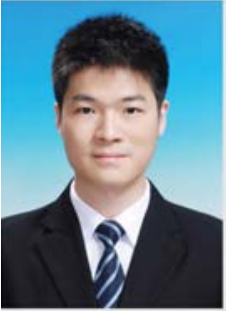}}]{Qi Wang}
(S'15-M'16) received the B.E. degree and the Ph.D. degree (with highest honor)
in electronic engineering from Tsinghua University, Beijing, China, in 2011
and 2016, respectively. From 2014 to 2015, he was a Visiting Scholar with the
Electrical Engineering Division, Centre for Photonic Systems, Department of
Engineering, University of Cambridge. From 2016 to 2017, he was a Research
Fellow with Southampton Wireless Group at the University of Southampton. Since 2017,
he has been a Senior Engineer at Huawei Technologies, China.

He has authored over 30 IEEE/OSA journal papers and several conference papers. He
also co-authored a book entitled \emph{Visible Light Communications: Modulation
and Signal Processing}, which was published by Wiley-IEEE Press and was selected by
the IEEE Series on Digital and Mobile Communication. He serves as
an Associate Editor for IEEE Access and a TPC member for many IEEE conferences, including
Globecom, GlobalSIP, CSNDSP, and IWCMC. His research interests include modulation and signal
processing for wireless communication and visible light communication. He was a recipient of
the Excellent Doctoral Dissertation of Chinese Institute of Electronics (CIE), Outstanding
Ph.D. Graduate of Tsinghua University, Excellent Doctoral Dissertation of Tsinghua University,
National Scholarship, and the Academic Star of Electronic Engineering Department in Tsinghua
University.
\end{IEEEbiography}

\end{document}